\documentclass[11pt]{article}
\usepackage[letterpaper,margin=1in]{geometry}
\usepackage[T1]{fontenc}
\usepackage{lmodern}
\usepackage{microtype}
\usepackage{amsmath,amssymb,amsthm,mathtools,bm}
\usepackage{booktabs,array,longtable,graphicx,enumitem,xcolor,needspace}
\usepackage{titlesec}
\usepackage{placeins}
\usepackage[labelsep=period]{caption}
\usepackage{tikz}
\usetikzlibrary{arrows.meta}
\usepackage[titles]{tocloft}
\usepackage[colorlinks=true,citecolor=blue,linkcolor=black,urlcolor=blue,bookmarks=true]{hyperref}
\usepackage[nameinlink,capitalize]{cleveref}
\hypersetup{
  pdftitle={EQP ⊆ IP Does Not Relativize},
  pdfauthor={Sinan Oral, Supartha Podder},
  pdfsubject={}
}
\setlist[itemize]{leftmargin=1.6em,itemsep=0.2em,topsep=0.35em}
\setlist[enumerate]{leftmargin=1.8em,itemsep=0.2em,topsep=0.35em}
\allowdisplaybreaks[2]
\numberwithin{equation}{section}
\titleformat{\section}{\normalfont\Large\bfseries}{\thesection}{1em}{}
\titleformat{\subsection}{\normalfont\large\bfseries}{\thesubsection}{1em}{}

\newtheorem{theorem}{Theorem}[section]
\newtheorem{lemma}[theorem]{Lemma}
\newtheorem{proposition}[theorem]{Proposition}

\newtheorem{external}[theorem]{External theorem}
\theoremstyle{definition}
\newtheorem{definition}[theorem]{Definition}
\theoremstyle{remark}

\newcommand{\E}{\mathbb E}
\newcommand{\Pp}{\mathbb P}
\newcommand{\R}{\mathbb R}
\newcommand{\N}{\mathbb N}
\newcommand{\one}{\mathbf 1}
\newcommand{\Unif}{\mathcal U}
\newcommand{\G}{\mathcal G}
\newcommand{\Pcal}{\mathcal P}
\newcommand{\Ecal}{\mathcal E}
\newcommand{\Good}{\mathsf{Good}}
\newcommand{\Nelig}{\mathcal N}
\newcommand{\poly}{\operatorname{poly}}
\newcommand{\sgn}{\operatorname{sgn}}
\newcommand{\forr}{\mathsf{forr}}
\newcommand{\diag}{\operatorname{diag}}
\newcommand{\osc}{\operatorname{osc}}
\newcommand{\DT}{\operatorname{DTdepth}}

\newcommand{\IP}{\ensuremath{\mathsf{IP}}}
\newcommand{\PH}{\ensuremath{\mathsf{PH}}}
\newcommand{\BQP}{\ensuremath{\mathsf{BQP}}}
\newcommand{\EQP}{\ensuremath{\mathsf{EQP}}}
\newcommand{\AM}{\ensuremath{\mathsf{AM}}}
\newcommand{\norm}[1]{\left\lVert #1\right\rVert}
\newcommand{\abs}[1]{\left|#1\right|}
\newcommand{\ip}[2]{\left\langle #1,#2\right\rangle}
\newcommand{\ket}[1]{\lvert #1\rangle}
\newcommand{\bra}[1]{\langle #1\rvert}

\newcommand{\Res}{\operatorname{Res}}
\newcommand{\Normal}{\mathrm N}

\newcommand{\errsm}{\varepsilon_{\rm sm}}

\title{$\mathsf{EQP}\subseteq\mathsf{IP}$ Does Not Relativize}
\author{
    \begin{tabular}{c@{\hspace{2.2em}}c} Sinan Oral & Supartha Podder \end{tabular}
    \\
    [0.9em]
    {\normalsize Department of Computer Science, Stony Brook University}\\[0.4em] {\normalsize \texttt{\char`\{soral, supartha\char`\}@cs.stonybrook.edu}}
}
\date{} 

\begin{document}
\maketitle
\vspace{-1.5em}
\begingroup

\let\small\normalsize
\begin{abstract}

We construct a recursive oracle $O$ and a unary language $L\in\EQP^O\setminus(\IP^O\cup\PH^O)$, strengthening the $\BQP$ versus $\IP$ oracle separation of Bouland et al.~\cite{BHNST} to exact quantum computation. The separation is based on a variant of the Forrelation problem in which the normalized correlation of one Boolean function with the Fourier transform of another is promised to be exactly $+1/2$~or~$-1/2$. The main technical contribution is a repair method that constructs distributions supported on these exact levels while preserving the indistinguishability needed for the $\IP$ and $\PH$ lower bounds. A uniform polynomial-time quantum algorithm distinguishes the two cases with certainty using only three oracle queries.

\end{abstract}
\endgroup

\clearpage
\begingroup
\normalsize
\setlength{\cftbeforesecskip}{0.35em}
\hypersetup{linkcolor=blue,linktoc=all}
\tableofcontents
\endgroup
\clearpage

\begingroup
\setlength{\parskip}{3pt plus 1pt minus 0.5pt}
\section{Introduction}\label{sec:introduction}

Can a classical polynomial-time verifier check a quantum computation by interacting with an unrestricted prover? Without an oracle, the answer is yes for polynomial-time quantum decision problems, since $\BQP\subseteq\mathsf{PSPACE}=\IP$, where the equality is Shamir's theorem~\cite{Shamir}. A relativizing proof of this containment would also work when the quantum algorithm, classical verifier, and prover have access to the same arbitrary Boolean oracle.

Bouland, Huang, Natarajan, Shalit, and Tal~\cite{BHNST} recently showed that this containment does not relativize by constructing an oracle $O$ for which $\BQP^O\not\subseteq\IP^O$. We show that this obstruction to relativization persists for $\EQP$, the class of decision problems solvable exactly in quantum polynomial time.

\begin{theorem}\label{thm:main}
There exist a recursive Boolean oracle $O$ and a unary language $L$ such that
\[
 L\in\EQP^O,\qquad L\notin\IP^O,\qquad L\notin\PH^O.
\]
\end{theorem}

The quantum algorithm makes at most three phase queries on each input and has a uniform, exact implementation over a fixed finite gate basis. The interactive-proof lower bound allows private verifier randomness, polynomially many messages, and unrestricted provers.

The separation uses Forrelation with promise values exactly $+1/2$ and $-1/2$. The main difficulty is to construct distributions supported on these levels that remain indistinguishable to classical tests. The same pair of distributions are used for both the $\IP$ and $\PH$ lower bounds.

\subsection{The exact Forrelation promise}

Let $f,g:\{0,1\}^n\to\{\pm1\}$, put $D=2^n$, and let $H$ be the normalized Fourier transform over $\mathbb F_2^n$. Their Forrelation is 
\[
 H_{xy}=D^{-1/2}(-1)^{x\cdot y},\qquad
 \forr(f,g):=D^{-1}f^THg=D^{-1}(Hf)^Tg.
\]
We distinguish the promised values $+1/2$ and $-1/2$, denoting the corresponding input sets by $\mathcal L_{n,+}$ and $\mathcal L_{n,-}$, with resources measured in the address length $n$.

The standard one-query interference test~\cite[full version, proof of Proposition~6]{AA} has a designated outcome of probability $p=(1+\forr(f,g))/2$. One amplitude-amplification iterate~\cite[Section~2]{BHMT} changes this probability to $p_{\mathrm{final}}=p(3-4p)^2$. For the two promised values,
\[
 \begin{aligned}
 \forr(f,g)=+\tfrac12&\quad\Longrightarrow\quad
 p=\tfrac34\quad\Longrightarrow\quad p_{\mathrm{final}}=0,\\
 \forr(f,g)=-\tfrac12&\quad\Longrightarrow\quad
 p=\tfrac14\quad\Longrightarrow\quad p_{\mathrm{final}}=1.
 \end{aligned}
\]
Accepting on the complementary outcome gives an exact three-query algorithm, whose circuit and exact gate implementation are given in Section~\ref{sec:quantum}.

Our classical lower bounds use hard distributions supported on the exact levels $\forr(f,g)=\pm1/2$. Concentration near these levels does not suffice, and conditioning a hard distribution on an exact level need not preserve indistinguishability.

\subsection{Technical overview}\label{subsec:technical-overview}

\paragraph{The comparison strategy.}
For each sign $\sigma\in\{+1,-1\}$, let $\Ecal_{n,\sigma}$ denote the exact law that we will construct. For each fixed polynomial bound on the classical resources and all sufficiently large eligible lengths, the goal is to show
\[
 \operatorname{supp}(\Ecal_{n,\sigma})\subseteq\mathcal L_{n,\sigma},
 \qquad
 \sup_C\left|\E_{\Ecal_{n,+}}C-\E_{\Ecal_{n,-}}C\right|=o(1),
\]
where $C$ ranges over tests induced by resource-bounded classical computations. For IP, it is the verifier's acceptance probability optimized over all prover strategies. After public-coin conversion~\cite{GS}, averaging over public randomness and maximizing over prover replies gives an average--maximum representation. This admits a convex extension whose Lipschitz constant is controlled by the verifier's resources. For PH, it is a bounded-depth Boolean circuit of size $2^{\poly(n)}$. Random restrictions reduce such circuits to decision trees, with a depth tail bound controlled by circuit depth and logarithmic size.

The distributions are independent of the tests and resource bounds, and the estimates are uniform over fixed oracle bits outside the two input truth tables. Theorem~\ref{thm:finite-common} gives the precise bound.

We compare both exact laws with the same uniform law $\Unif_n$. An $o(1)$ bound for each comparison gives the desired positive--negative gap by the triangle inequality. The comparisons pass through a conditional product law $\Pcal_{n,\sigma}$ and a Gaussian-sign law $\G_{n,\sigma}$, as shown below.

\par\noindent\begin{minipage}{\linewidth}
\centering
\hypersetup{linkcolor=black}
\vspace{0.5\baselineskip}
\begin{tikzpicture}[
  comparison law/.style={draw=black,thin,font=\large,inner sep=5pt},
  comparison label/.style={font=\normalsize,inner sep=2pt},
  >={Computer Modern Rightarrow}
]
\node[comparison law] (exact) at (0,0) {$\Ecal_{n,\sigma}$};
\node[comparison law] (product) at (4.8,0) {$\Pcal_{n,\sigma}$};
\node[comparison law] (gaussian) at (9.6,0) {$\G_{n,\sigma}$};
\node[comparison law] (uniform) at (14.4,0) {$\Unif_n$};
\draw[<->,thin,shorten <=4pt,shorten >=4pt] (exact.east) --
  node[comparison label,above=6pt] {Sec.~\ref{sec:coupling}}
  node[comparison label,below=11pt] {count repair} (product.west);
\draw[<->,thin,shorten <=4pt,shorten >=4pt] (product.east) --
  node[comparison label,above=6pt] {Secs.~\ref{sec:ip}--\ref{sec:ph}}
  node[comparison label,below=11pt] {magnitude replacement} (gaussian.west);
\draw[<->,thin,shorten <=4pt,shorten >=4pt] (gaussian.east) --
  node[comparison label,above=6pt] {Secs.~\ref{sec:ip}--\ref{sec:ph}}
  node[comparison label,below=11pt] {Gaussian comparison} (uniform.west);
\end{tikzpicture}
\par\vspace{0.5\baselineskip}
\raggedright
\noindent Each arrow compares acceptance expectations for the $\IP$ and $\PH$ test families. 
\par
\end{minipage}\par\addvspace{0.5\baselineskip}

\paragraph{From a Gaussian reference to conditional independence.}
We construct the exact distributions from the Gaussian reference in two steps. We first obtain conditional independence and then repair the sign counts to enforce the promise. It suffices to describe the positive construction, since negating $g$ gives the negative one. A natural starting point is a noisy variant of the Gaussian-sign Forrelation distribution~\cite{Aaronson,AA,BHNST}, given by
\[
 \G_{n,+}:\qquad f=\sgn X,\qquad
 g=\sgn(\rho HX+sY),\qquad s^2=1-\rho^2,
\]
where $X,Y$ are independent standard Gaussian vectors, $0<\rho<1$, and signs are taken coordinatewise. The normalized Fourier pairing of $X$ and $\rho HX+sY$ has expectation $\rho$, although every entry of their cross-covariance block has magnitude only $\rho/\sqrt D$. The parameter $\rho$ calibrates the Forrelation, while the independent noise keeps the joint covariance nonsingular for the analytic comparisons.

Count repair requires independence conditional on $f$. The coordinates of $g$ are independent given $X$, but generally not given $f=\sgn X$, since $H$ mixes the remaining random magnitudes $|X_i|$. Writing $X_i=f_i|X_i|$, we replace the centered magnitude fluctuations by independent Gaussians with matching first two moments. The mean contributes a multiple of $Hf$, and orthogonality makes the transformed Gaussian noise independent across coordinates given $f$. Adding $sY$ gives
\[
 \Pcal_{n,+}:\qquad f\text{ uniform},\qquad
 g=\sgn(aHf+bZ),
\]
with fixed positive constants $a,b$ and an independent standard Gaussian vector $Z$. Conditional on $f$, the coordinates of $g$ are independent, and equal values of $(Hf)_i$ give equal probabilities of $g_i=+1$.

\paragraph{Exactness through counts, randomness through positions.}
Fix $f$ and let $u=\sqrt D\,Hf\in\mathbb{Z}^D$. The positive promise can then be written as the integer constraint
\[
S(f,g):=\sum_i u_i g_i=\frac{D^{3/2}}2.
\]
A \emph{weight bin} consists of coordinates with the same coefficient $u_i=w$. If a bin contains $N$ coordinates, of which $K$ have positive signs, its contribution to $S(f,g)$ is $w(2K-N)$, regardless of their positions. Under $\Pcal_{n,+}$, conditional on $f$ and $K$, all $K$-element sets of positive positions within the bin are equally likely. We can therefore prescribe counts to enforce the exact equality while keeping the positions random.

Section~\ref{sec:laws} selects small positive weights that permit an exact integer correction and proves that their bins have enough capacity. We calibrate the correlation parameter so that a Gaussian approximation to the expected Forrelation equals $1/2$, then bound the remaining discrepancy. After sampling $f$ and the signs outside the selected bins, we round and repair the expected counts and sample the positive positions uniformly within each bin. Each count changes by a vanishing fraction of its bin size. Exceptional outcomes are replaced by a fixed promised input, ensuring exact support with a small additional comparison error.

\paragraph{Why the repair is difficult to detect.}
A small fractional change in a bin's positive count does not imply closeness in total variation, since the change may exceed the usual binomial fluctuations. A direct Lipschitz comparison is also inadequate, since changing even one sign gives infinity-norm distance two. Instead, following the product-to-slice coupling of Filmus--Mossel~\cite{FM}, we use a shared random ordering within each bin. The reference sample $X$ assigns positive signs to the first $\ell$ positions, where $\ell$ is binomial, while the exact-count sample $Y$ uses the first $K$, its prescribed count. Their positive sets are nested and differ in exactly $|\ell-K|$ positions, which are spread uniformly by the random ordering.

For convex tests, fix $f$ and outside signs for which repair is feasible. Given either coupled sample, the conditional mean of the other is close to it in expected infinity norm. Conditional Jensen's inequality and the Lipschitz bound then give a two-sided comparison of test expectations. Bounding the maximum discrepancy across bins avoids a separate error from each bin.

For Boolean tests, we fix the same feasible outer data and realize the ordering by independent uniform variables. Leaving a narrow window around each bin's threshold undecided gives an independent random restriction $\eta$ that $X$ always extends and $Y$ extends with high probability. Therefore,
\[
 \Pp[C(X)\ne C(Y)]
 \le \Pp[C|_\eta\text{ is nonconstant}]
      +\Pp[Y\text{ does not extend }\eta].
\]
Restriction bounds and concentration control the two terms. The independence of $\eta$ is essential, since the actual disagreement set need not give a product restriction.

\paragraph{Completing the comparisons with uniform inputs.}
The remaining two comparisons are analytic. For magnitude replacement, matching the first two moments cancels the quadratic Taylor polynomial, leaving a third-derivative remainder. Gaussian comparison uses second-derivative estimates to control the effect of the small cross-covariances. The small entries of $H$ yield vanishing errors once the relevant sums of derivative contributions are controlled. Section~\ref{sec:shared-analysis} develops the shared tools using the Lindeberg method~\cite{Chatterjee} and Price's formula~\cite{Price}. Applying these tools to the two classical models requires different ways of handling the discontinuity of the sign map.

\emph{Interactive proofs.} Softmax smoothing of the verifier's average--maximum acceptance function~\cite{AGH,BHNST} gives a convex approximation with resource-controlled derivatives, but its composition with the sign map remains discontinuous. We use the Gaussian-sign comparison of~\cite{BHNST} for the covariance step. For magnitude replacement, we regularize the sign map near zero, bound the third derivatives, and control the error from removing this regularization at both endpoints.

\emph{The polynomial hierarchy.} We pass to the test's multilinear extension and use Rossman's restriction bounds~\cite[Theorem~3 and Section~1.2]{Rossman19} to control its weighted derivatives. Averaging over the independent Gaussian noise already present in the construction then yields a smooth function with exactly the same acceptance expectation. This supplies the derivative bounds for covariance interpolation and magnitude replacement without an additional sign-approximation error.

\paragraph{From indistinguishability to an oracle separation.}
Indistinguishability guarantees a failing promised input for every fixed classical candidate at all sufficiently large eligible lengths. Since the promise sets and finite acceptance games are decidable, exhaustive search finds such an input without sampling the hard distributions. Section~\ref{sec:oracle} fixes a sufficiently long oracle prefix at each stage to preserve earlier failures, yielding a recursive oracle and a unary language decided by the exact quantum algorithm.

\subsection{Relation to previous work}

Our $\IP$ lower bound builds on the convex-analytic framework of~\cite{BHNST}. Those estimates handle Gaussian-sign reference distributions, while our exact separation additionally requires constructing distributions supported on the exact promise sets and proving that count repair and magnitude replacement preserve the required indistinguishability.

Mantri~\cite[Theorems~1.1 and~1.3]{Mantri} recently constructed an oracle language in $(\EQP^O\cap\IP^O)\setminus\PH^O$, with a public-coin protocol, perfect completeness, and an efficient quantum honest prover. His exact algorithm uses $3^{h(n)}$ queries for an unbounded recursion height $h(n)$, although the query count can grow arbitrarily slowly~\cite[Corollary~1.5]{Mantri}. Our language lies outside both $\IP^O$ and $\PH^O$, yet admits an exact algorithm making only three queries. Both lower bounds follow from the same pair of exact-promise distributions.

Exact quantum advantages include the Deutsch--Jozsa query separation against deterministic classical algorithms~\cite{DJ}, the oracle separations of Berthiaume and Brassard~\cite{BB}, and the exact algorithm for Simon's problem of Brassard and H\o yer~\cite{BH}. Green and Pruim~\cite{GP} separated $\EQP^O$ from $\mathsf{P}^{\mathsf{NP}^O}$.

Forrelation was introduced by Aaronson~\cite{Aaronson} as a candidate for an oracle separation between $\BQP$ and $\PH$. Aaronson--Ambainis~\cite{AA} proved an essentially optimal separation between one quantum query and randomized query complexity, and Raz--Tal~\cite{RT} used a Forrelation variant to separate $\BQP$ from $\PH$. Bansal--Sinha~\cite{BansalSinha} established near-optimal randomized query lower bounds for $k$-Forrelation using Gaussian interpolation and integration by parts.

Girish--Servedio~\cite[Theorem~4]{GirishServedio} proved a randomized query lower bound of $\widetilde\Omega(D^{1/4})$ for the extremal promise $\forr(f,g)\in\{\pm1\}$, which admits an exact one-query algorithm. Canonne, Chen, and Mestre~\cite[Theorem~3]{CCM} subsequently proved a nonadaptive randomized query lower bound of $\Omega(D^{1/2-o(1)})$ for the same promise. Our promise instead uses $\{\pm1/2\}$ and admits an exact three-query algorithm, while the corresponding hard distributions establish lower bounds against interactive proofs and $\PH$.

\paragraph{Organization.}
Sections~\ref{sec:overview} and~\ref{sec:quantum} state the promise problem and give its exact quantum solver. Section~\ref{sec:laws} constructs the distributions, and Section~\ref{sec:coupling} proves stability under count repair. Section~\ref{sec:shared-analysis} supplies the analytic comparison tools, and Sections~\ref{sec:ip} and~\ref{sec:ph} apply them to the two classical models. Section~\ref{sec:oracle} constructs the recursive oracle.

\par
\endgroup
\section{Preliminaries}\label{sec:preliminaries}\label{sec:overview}
We specify oracle access and the classical test families, then state the indistinguishability theorem. The quantum running time is polynomial in the address length $n$, which later becomes the length of the unary input $1^n$.

\subsection{Oracle access and the exact gate basis}\label{subsec:models}
An oracle is a total Boolean function $O:\{0,1\}^*\to\{0,1\}$. A classical query on $z$ returns the bit $O(z)$. At address length $n$, the \emph{target block} consists of all strings $0x,1x$ with $x\in\{0,1\}^n$, encoding the truth tables of $f$ and $g$ via
\[
 f(x)=(-1)^{O(0x)},\qquad g(x)=(-1)^{O(1x)}.
\]
Thus each string in this block has length $n+1$. All oracle bits outside the block are called \emph{side-oracle bits}.

For each query length $m\ge1$, quantum access is the \emph{phase query} on $(\mathbb C^2)^{\otimes m}$ defined by
\[
 U_{O,m}\ket z=(-1)^{O(z)}\ket z,
 \qquad z\in\{0,1\}^m.
\]
The query acts as the identity on all other registers. This unitary satisfies $U_{O,m}^\dagger=U_{O,m}^{-1}=U_{O,m}$. Each application counts as one quantum query. We write $U_O$ when the query length is clear. We use the fixed exact quantum gate basis $\{H_2,T,\mathrm{CNOT}\}$, where
\[
 H_2=\frac1{\sqrt2}\begin{pmatrix}1&1\\1&-1\end{pmatrix},
 \qquad T=\diag(1,e^{i\pi/4}).
\]
Section~\ref{sec:quantum} implements all non-oracle gates exactly, and Appendix~\ref{app:real} gives a realization using Hadamard and reversible classical gates. The usual Boolean oracle gate $\ket z\ket a\mapsto\ket z\ket{a\oplus O(z)}$ also provides phase access. Preparing the target in $\ket{-}=(\ket0-\ket1)/\sqrt2$ gives
\[
 \ket z\ket{-}\longmapsto(-1)^{O(z)}\ket z\ket{-}.
\]
The target is prepared exactly from the fixed basis, so each phase query uses one Boolean oracle query.

In an interactive proof, the prover has unrestricted computation and oracle access, while the verifier has polynomial resources. We allow private verifier randomness unless a public-coin protocol is specified. On a fixed input, a verifier or predicate with bounded running time can query only finitely many side-oracle bits across all possible computation paths. Fixing these bits lets us view its computation as a test of $f$ and $g$. Our bounds hold uniformly over their assignments, with the verifier or predicate still accessing them through classical queries.

The oracle is \emph{recursive} if a deterministic Turing machine computes every bit and halts on every input. No running-time bound is required. In contrast, the quantum algorithm must be uniform and run in polynomial time on every input.

\subsection{Fourier notation and the exact Forrelation promise}\label{subsec:notation}
Put $D=2^n$. All operations on elements of $\mathbb F_2^n$ are over $\mathbb F_2$, while vector inner products and expectations are taken over the reals. All logarithms are natural unless a base is displayed. The normalized Fourier transform over $\mathbb F_2^n$ is represented by the Hadamard matrix
\[
 H=H_D,\qquad H_{xy}=D^{-1/2}(-1)^{x\cdot y},\qquad
 H=H^T=H^{-1}.
\]
For $f,g\in\{\pm1\}^D$, their Forrelation is
\begin{equation}\label{eq:phi}
 \begin{aligned}
 \forr(f,g)&=D^{-1}f^THg=D^{-1}(Hf)^Tg\\
 &=D^{-3/2}\sum_{x,y}f(x)(-1)^{x\cdot y}g(y).
 \end{aligned}
\end{equation}
By Cauchy--Schwarz, $\forr(f,g)\in[-1,1]$. The Fourier coefficients of $f$ will index the bins in Section~\ref{sec:laws}.

We use address lengths in the polynomial-time recognizable infinite set
\[
 \Nelig=\{4,8,12,\ldots\} = 4\mathbb N_{>0}.
\]
These lengths make the target in the integer constraint below integral, provide the divisibility used in Lemma~\ref{lem:counts}, and allow the explicit default inputs of Section~\ref{sec:quantum}. This is a convenient sufficient choice. The final language rejects lengths outside $\Nelig$. For $n\in\Nelig$, the positive and negative promise sets are
\[
 \mathcal L_{n,+}=\{(f,g):\forr(f,g)=1/2\},\qquad
 \mathcal L_{n,-}=\{(f,g):\forr(f,g)=-1/2\}.
\]
A pair $(f,g)$ is a \emph{promised input} if it belongs to either set. We use $\sigma\in\{+1,-1\}$ for the sign and abbreviate $+1,-1$ by $+, -$ in subscripts.

To check membership exactly, write $W=\sqrt D\,H$, whose entries are $\pm1$. Then
\[
 (f,g)\in\mathcal L_{n,\sigma}
 \quad\Longleftrightarrow\quad
 f^TWg=\sigma\,\frac{D\sqrt D}{2}.
\]
This equivalence follows by multiplying the Forrelation equality by $D\sqrt D$. Both sides are integers for $n\in\Nelig$, so each promise set is finite and uniformly decidable, as needed for the exhaustive oracle construction. We write $\Unif_n$ for the uniform distribution on $\{\pm1\}^{2D}$.

\subsection{Classical test families}\label{subsec:test-models}
We compare expectations for two families of functions of $f$ and $g$. Section~\ref{sec:ip} represents interactive verifiers by convex average--maximum functions, and Section~\ref{sec:ph} represents PH predicates by bounded-depth Boolean circuits. The tests below have $m$ inputs, with $m=2D$ for the pair $(f,g)$.

\paragraph{Average--maximum circuits.}
After the public-coin reduction in Section~\ref{subsec:full-ip}, a verifier's optimal acceptance probability is evaluated by averaging over fresh public randomness and maximizing over prover replies. This motivates the following class.
\begin{definition}\label{def:avgmax}
A $(d,B,\ell)$ average--maximum circuit on $m$ real inputs is a rooted tree of depth at most $d$ and fanin at most $2^B$, with uniform-average and maximum gates and affine leaves $c+\ip ux$ satisfying $\norm u_1\le\ell$. Thus $B$ bounds the base-two logarithm of the fanin, even though unqualified logarithms elsewhere are natural. It defines a convex function on $\R^m$. We call it an acceptance function if its values on sign vectors lie in $[0,1]$.
\end{definition}
Affine leaves are $\ell$-Lipschitz in the infinity norm, and averages and maxima preserve convexity and this Lipschitz bound. Section~\ref{subsec:full-ip} bounds the circuit parameters in terms of verifier resources.

\paragraph{Boolean circuits and restrictions.}
Boolean circuits use unbounded-fanin AND/OR gates, with negations at the inputs. Depth counts gate layers, total size includes input nodes, and a formula is a circuit with a tree structure. Our indistinguishability bound uses the following property of small bounded-depth circuits. Even after any prior deterministic restriction, randomly fixing most remaining inputs is unlikely to leave a function requiring many further queries.

A \emph{restriction} fixes some coordinates to signs and leaves the others free. Let $\Res_t$ independently leave each coordinate free with probability $t$, fixing it to a fair sign otherwise. Write $\DT(C)$ for the minimum worst-case decision-tree depth of $C$, and $C|_{\tau,\eta}$ for first imposing $\tau$, then applying $\eta$ to the remaining coordinates.
\begin{definition}\label{def:criticality}
A Boolean function $C:\{\pm1\}^m\to\{0,1\}$ satisfies a restriction-stable criticality bound $\Lambda\ge1$ through order three if, for every deterministic restriction $\tau$, every $0\le t\le1$, and $r=1,2,3$,
\begin{equation}\label{eq:criticality}
 \Pp_{\eta\sim \Res_t}\bigl[\DT(C|_{\tau,\eta})\ge r\bigr]
 \le(t\Lambda)^r.
\end{equation}
\end{definition}
When $t\Lambda$ is small, the case $r=1$ ensures that the restricted function is constant with high probability. Section~\ref{sec:coupling} uses this case to control repair, while the higher orders control the analytic comparisons. Section~\ref{subsec:ph-criticality} proves the condition for each fixed circuit depth with $\Lambda$ polynomial in logarithmic size.

\subsection{Classical indistinguishability}\label{subsec:finite-bound}
Section~\ref{sec:laws} constructs distributions $\Ecal_{n,\sigma}$ supported on $\mathcal L_{n,\sigma}$ and depending only on $n,\sigma$. The following bound applies to both test families.
\begin{theorem}[Classical indistinguishability]\label{thm:finite-common}
Let $q(n)\ge n+1$ be a fixed polynomial. For all sufficiently large $n\in\Nelig$, the distributions $\Ecal_{n,+}$ and $\Ecal_{n,-}$ constructed in Section~\ref{sec:laws} satisfy
\begin{equation}\label{eq:finite-common}
 \sup_C\abs{\E_{\Ecal_{n,+}}C-\E_{\Ecal_{n,-}}C}
 \le O\!\left(q(n)^2n^{1/4}2^{-n/12}\right),
\end{equation}
where $C$ ranges over either of the following families.
\begin{enumerate}[label=(\roman*)]
\item Average--maximum acceptance functions with depth, log fanin, and affine-leaf gradient $\ell_1$ norm at most $q(n)$.
\item Boolean functions with restriction-stable criticality bound $\Lambda\le q(n)$ through order three, as in Definition~\ref{def:criticality}.
\end{enumerate}
The implicit constant is absolute; the threshold on $n$ may depend on $q$. The bound holds uniformly over all fixed assignments of side-oracle bits.
\end{theorem}
Lemma~\ref{lem:critical-envelope} places every fixed-depth circuit family with polynomially bounded log size in family (ii), for a suitable polynomial $q$.

For a fixed side assignment, a test that is at least $2/3$ on every positive promised input and at most $1/3$ on every negative promised input would satisfy
\[
 \E_{\Ecal_{n,+}}C-\E_{\Ecal_{n,-}}C\ge1/3.
\]
For each fixed $q$, the bound in \eqref{eq:finite-common} tends to zero, so every allowed test must eventually fail on some promised input. Section~\ref{sec:oracle} finds such an input by exhaustive search over the decidable promise sets. The distributions establish its existence; the search need not sample them or compute their probabilities.

\section{Exact quantum algorithm and default inputs}\label{sec:quantum}

The standard Forrelation interference test, followed by one amplitude-amplification step, distinguishes $\mathcal L_{n,+}$ from $\mathcal L_{n,-}$ exactly. The resulting algorithm uses three phase queries and gates from the basis of Section~\ref{subsec:models}. We also construct a uniformly computable default input in each promise set for the sampling procedure and the oracle construction.

\subsection{The exact three-query algorithm}
\begin{lemma}[One-query Forrelation test~{\cite[proof of Proposition~6, $k=2$]{AA}}]\label{lem:interference}
For $D=2^n$, there is a quantum test $A$ such that
\begin{equation}\label{eq:quantum-p}
 \Pr\!\left[A^{f,g}(1^n)=1\right]
   =\frac{1+\forr(f,g)}2
\end{equation}
for every pair $f,g\in\{\pm1\}^D$. The test uses one tagged phase query and $O(n)$ non-oracle gates.
\end{lemma}

\begin{figure}[htbp]
\centering
\begin{tikzpicture}[
  x=1.15cm,y=1cm,
  gate/.style={draw,fill=white,minimum height=0.53cm,
               minimum width=0.63cm,inner sep=3pt},
  every node/.style={font=\normalsize},
  line width=0.55pt
]
  \draw[densely dashed,rounded corners=2pt,black!50]
    (0.05,0.65) rectangle (7.10,-1.75);
  \node[fill=white,inner sep=3pt] at (3.58,0.65)
    {$A$ (before measurement)};

  \foreach \y in {0,-1.1} {
    \draw (0,\y) -- (8.75,\y);
  }
  \node[anchor=east] at (-0.12,0) {$b:\ket0$};
  \node[anchor=east] at (-0.12,-1.1) {$x:\ket{0^n}$};

  \node[gate] at (1.15,0) {$H_2$};
  \node[gate] at (1.15,-1.1) {$H_2^{\otimes n}$};

  \node[draw,fill=white,minimum width=0.94cm,
        minimum height=1.65cm,align=center] at (3.0,-0.55)
    {$U_O$};

  \draw (5.0,0) -- (5.0,-1.1);
  \fill (5.0,0) circle (2pt);
  \node[gate] at (5.0,-1.1) {$H_2^{\otimes n}$};
  \node[gate] at (6.35,0) {$H_2$};

  \node[gate,minimum width=0.70cm] at (8.0,0) {};
  \draw (7.78,-0.10) arc[start angle=180,end angle=0,radius=0.22];
  \draw[->] (8.0,-0.10) -- (8.15,0.16);
\end{tikzpicture}
\caption{The preliminary one-query Forrelation interference circuit, equivalent to the $k=2$ construction of Aaronson--Ambainis~\cite[proof of Proposition~6]{AA}. The branch qubit $b$ serves as the oracle tag. Before amplitude amplification, the test outputs $1$ when $b=0$ is measured; the exact solver accepts the complementary outcome after amplification. Ancillas, initialized and returned to zero, are omitted.}
\label{fig:one-query}
\end{figure}
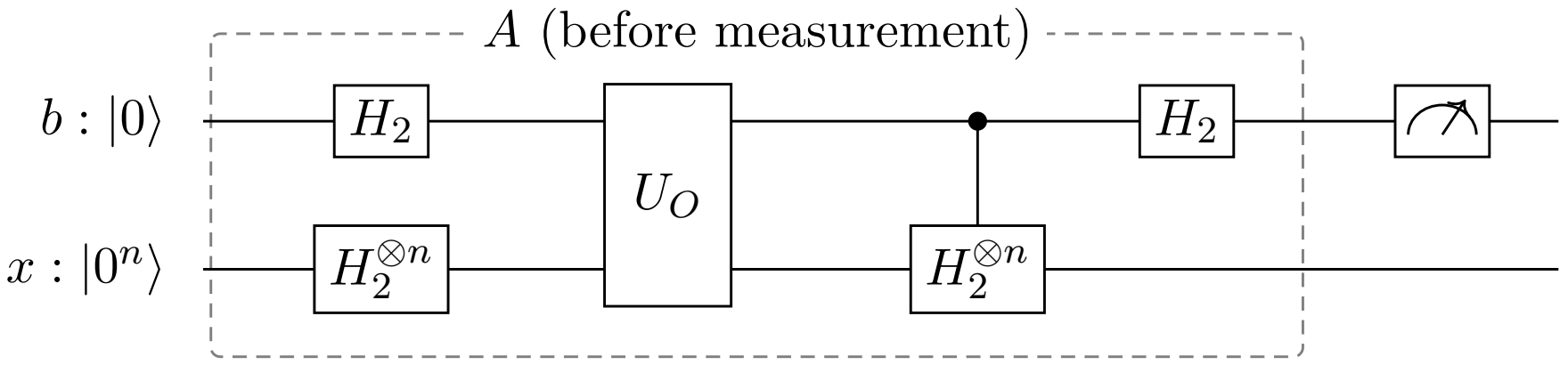

\begin{proof}
For a sign function $h$, write $\ket{u_h}=D^{-1/2}\sum_x h(x)\ket x$. Figure~\ref{fig:one-query} shows the circuit, whose branch qubit $b$ also serves as the oracle tag. The initial Hadamards prepare
\[
 \frac{\ket0+\ket1}{\sqrt2}\otimes\frac1{\sqrt D}\sum_x\ket x.
\]
The tagged query applies the phase $f(x)$ on branch zero and $g(x)$ on branch one. The controlled $H_D$ then gives the state
\[
 \frac{\ket0\ket{u_f}+\ket1 H_D\ket{u_g}}{\sqrt2}.
\]
Since both branch states have unit norm and real amplitudes, the final branch Hadamard gives
\[
 \Pr[b=0]
 =\frac14\norm{\ket{u_f}+H_D\ket{u_g}}_2^2
 =\frac{1+D^{-1}f^TH_Dg}{2}
 =\frac{1+\forr(f,g)}2.
\]
The test outputs $1$ on outcome $b=0$. It uses $O(n)$ Hadamards and controlled Hadamards, whose exact implementation in the fixed basis is given below.
\end{proof}

\begin{proposition}[Exact promise solver]\label{prop:exact-solver}
There is a uniform polynomial-time quantum algorithm over the fixed gate basis of Section~\ref{subsec:models} that makes three phase queries and, with certainty, accepts $\mathcal L_{n,+}$ and rejects $\mathcal L_{n,-}$.
\end{proposition}

\begin{proof}
Let $A$ denote the unitary part of the one-query test, including preparation and clean ancillas, acting on $r$ qubits initialized to zero.

\smallskip\noindent\textbf{Exact amplification.}
Let $\Pi$ project onto branch $b=0$, and write
\[
 \ket\psi=A\ket{0^r}=\ket{\psi_0}+\ket{\psi_1},\qquad
 \ket{\psi_0}=\Pi\ket\psi,\quad
 \ket{\psi_1}=(I-\Pi)\ket\psi.
\]
By \eqref{eq:quantum-p}, $p=\|\ket{\psi_0}\|^2$ is $3/4$ on the positive promise and $1/4$ on the negative promise. One amplitude-amplification step~\cite[Section~2]{BHMT} uses the reflections
\[
 S_{\rm mark}=I-2\Pi,\qquad
 S_0=I-2\ket{0^r}\bra{0^r}.
\]
Since $AS_0A^\dagger=I-2\ket\psi\bra\psi$, the iterate $Q=AS_0A^\dagger S_{\rm mark}$ and its final branch-zero probability $p_{\rm final}$ satisfy
\begin{align}
 Q\ket\psi
  &=(4p-3)\ket{\psi_0}+(4p-1)\ket{\psi_1},\label{eq:amp-state}\\
 p_{\rm final}&=p(4p-3)^2.\label{eq:amp-prob}
\end{align}
The final probability of $b=0$ is therefore $0$ for $p=3/4$ and $1$ for $p=1/4$. Accepting on $b=1$ decides the promise exactly, using the same fixed circuit for both promise values.

\smallskip\noindent\textbf{Query complexity.}
The full circuit $QA=AS_0A^\dagger S_{\rm mark}A$ contains two copies of $A$ and one of $A^\dagger$. We implement $A^\dagger$ by reversing the gate order and inverting each gate. The phase query $U_O$, $H_2$, and CNOT are self-inverse, and $T^{-1}=T^7$. Thus $A^\dagger$ also uses one oracle query and polynomially many gates. The reflections use no queries, giving three in total. We now verify the exact implementation of the non-oracle gates.

\smallskip\noindent\textbf{Exact gate synthesis.}
The controlled $H_D=H_2^{\otimes n}$ is a product of $n$ controlled single-qubit Hadamards, and the all-zero reflection uses reversible conjunctions. A controlled Hadamard has matrix $\diag(I_2,H_2)$, and Toffoli is a permutation matrix, so both have entries in $\mathbb Z[1/\sqrt2,i]$. By Giles--Selinger~\cite[Theorem~1]{GilesSelinger}, each has an exact Clifford+$T$ synthesis with ancillas initialized and returned to zero. Since these are fixed gates of constant arity, we may fix constant-size syntheses once and for all.

The Clifford generators reduce to our basis using $S=T^2$, $Z=T^4$, and $X=H_2ZH_2$. The scalar generator $\omega_8=e^{i\pi/4}$ can be implemented on a clean ancilla by $XTX$, which maps $\ket0$ to $\omega_8\ket0$.

\smallskip\noindent\textbf{Reflections and uniformity.}
The reflection $S_{\rm mark}$ is $XZX$ on the branch bit. The reflection $S_0$ acts on all $r$ qubits of $A$, including its synthesis ancillas. We implement it by flipping these qubits with $X$, computing their AND into additional clean work bits by a Toffoli chain, applying $Z$ to the final AND bit, and uncomputing the chain and the $X$ gates. This changes the phase of precisely the original all-zero basis state. The additional work bits and any ancillas used to synthesize the reflection gates are separate from the $r$ qubits on which $S_0$ acts, and are initialized and returned to zero.

The fixed syntheses and reflection construction give a deterministic polynomial-time procedure for generating the circuit, with gate count and workspace polynomial in $n$.
\end{proof}

Appendix~\ref{app:real} gives an alternative three-query implementation using Hadamard and reversible classical gates.

\subsection{Computable default inputs}
For each eligible length, we give a fixed pair in each promise set, called a \emph{default input}, with entries computable in $O(n)$ time. The construction uses a sign function $f_\star$ satisfying $Hf_\star=f_\star$. Its Forrelation with $g$ is their normalized inner product, so changing its sign on one quarter of the addresses gives a function $g$ with Forrelation $1/2$.

Since eligible lengths are even, write $x=(x_L,x_R)$ with $n/2$ bits in each half, and define
\[
 f_\star(x)=(-1)^{x_L\cdot x_R}.
\]
For an address $y=(y_L,y_R)$,
\[
 (Hf_\star)(y)=D^{-1/2}
   \sum_{x_L,x_R}(-1)^{x_L\cdot x_R+y_L\cdot x_L+y_R\cdot x_R}
   =(-1)^{y_L\cdot y_R}=f_\star(y).
\]
Indeed, $\sum_{x_R}(-1)^{(x_L+y_R)\cdot x_R}$ equals $\sqrt D$ if $x_L=y_R$ and zero otherwise, and the remaining factor is $(-1)^{y_L\cdot y_R}$.

Let $\mathcal B_\star$ be the set of addresses whose first two bits are zero. Every eligible length is at least two, so $|\mathcal B_\star|=D/4$. Define $g_\star=f_\star$ off $\mathcal B_\star$ and $g_\star=-f_\star$ on $\mathcal B_\star$. Then $(f_\star,g_\star)$ is a positive default input, since
\begin{equation}\label{eq:filler}
 \forr(f_\star,g_\star)=D^{-1}(Hf_\star)^Tg_\star
          =D^{-1}f_\star^Tg_\star
          =\frac{D-2|\mathcal B_\star|}{D}=\frac12.
\end{equation}
The pair $(f_\star,-g_\star)$ belongs to $\mathcal L_{n,-}$. Both pairs are uniformly computable, since each function value can be evaluated in $O(n)$ time from an eligible length $n$ and an address $x$. Section~\ref{sec:laws} uses the positive pair as the default sampler output, and negating its second function gives the negative default. The oracle construction in Section~\ref{sec:oracle} uses the positive pair for blocks not otherwise selected, including the finitely many eligible lengths before the distributional estimates apply.

\section{\texorpdfstring{Constructing the exact distributions $\Ecal_{n,\sigma}$}{Constructing the exact distributions E(n,sigma)}}\label{sec:laws}

We construct distributions $\Ecal_{n,\sigma}$ supported on the promise sets $\mathcal L_{n,\sigma}$ without reference to the classical test. Their indistinguishability is proved in Sections~\ref{sec:coupling}--\ref{sec:ph}. Recall $W=\sqrt D\,H$ and $u :=Wf$. We define the \emph{energy} as the integer-valued quantity
\[
 S(f,g):=\sum_i u_i g_i=D^{3/2}\forr(f,g).
\] 
For the positive distribution, the target energy is $S_*:=D^{3/2}/2$. A \emph{weight bin} $I_w(f)=\{i:u_i=w\}$ of size $N$ contributes $w(2K-N)$ to the energy when $g$ has exactly $K$ plus signs in that bin. Since this contribution does not depend on their positions, we first choose the counts to attain $S_*$ and then sample the positions of the plus signs uniformly within each bin. Negating $g$ reverses the sign of the energy and gives the negative distribution.

Section~\ref{subsec:reference-distributions} introduces the reference distributions. Sections~\ref{subsec:weight-bins} and~\ref{subsec:energy-calibration} bound bin sizes and the conditional-mean energy discrepancy. These estimates yield feasible count corrections in Section~\ref{subsec:count-repair} and the sampling rule in Section~\ref{subsec:exact-sampling}.

\subsection{\texorpdfstring{The distributions $\G_{n,\sigma}$ and $\Pcal_{n,\sigma}$}{The distributions G(n,sigma) and P(n,sigma)}}\label{subsec:reference-distributions}

We begin with a noisy, tunable variant of Gaussian-sign Forrelation~\cite{Aaronson}. The correlation parameter lets us center the energy near $S_*$, while independent noise keeps the Gaussian covariance nonsingular for the later comparisons. Write $\Normal(\mu,\Sigma)$ for a Gaussian distribution, and take signs coordinatewise with $\sgn(0)=+1$. For now let $0<\rho<1$, put $s=\sqrt{1-\rho^2}$, and draw independent $X,Y\sim\Normal(0,I_D)$. For $\sigma\in\{+1,-1\}$, define
\[
 \G_{n,\sigma}:\quad f=\sgn X,\qquad
       g=\sgn(\sigma\rho HX+sY).
\]
The Gaussian vectors before taking signs are individually standard normal. The second has covariance $\rho^2HH^T+s^2I=(\rho^2+s^2)I=I$. Independence of $X,Y$ gives cross-covariance $\E[X(\sigma\rho HX+sY)^T]=\sigma\rho H^T=\sigma\rho H$. Their joint covariance is
\begin{equation}\label{eq:covariance}
 \Sigma_{n,\sigma}=\begin{pmatrix}I&\sigma\rho H\\
                       \sigma\rho H&I\end{pmatrix}.
\end{equation}
Every entry of the cross-covariance block has magnitude $\rho/\sqrt D$, although the Fourier pairing has constant mean
\[
 \E[D^{-1}X^TH(\sigma\rho HX+sY)]
 =\sigma\rho D^{-1}\E\|X\|_2^2=\sigma\rho.
\]
Here $H^2=I$ gives the first term and independence makes the $X,Y$ term vanish. For any $v\in\R^D$, the block vectors $(v,\pm\sigma Hv)$ have eigenvalues $1\pm\rho$ under $\Sigma_{n,\sigma}$. Together they span $\R^{2D}$. The later comparisons exploit both the small cross-covariances and the positive lower bound $1-\rho$.

\paragraph{\texorpdfstring{Obtaining the conditional product distribution $\Pcal_{n,\sigma}$.}{Obtaining the conditional product distribution P(n,sigma).}}
In the repair, we condition on $f$. Under $\G_{n,\sigma}$, the coordinates of $g$ are independent given $X$, but conditioning only on $f=\sgn X$ leaves the magnitudes $|X_i|$ random. The transform $H$ mixes these hidden magnitudes, so the signs of $g$ need not be independent conditional on $f$. We obtain conditional independence by replacing the centered magnitudes with Gaussian variables having the same variance.

Write $X_i=f_iR_i$, where $R_i=|X_i|$ are independent half-normal variables, independent of the uniform sign vector $f$, with mean $m_0=\sqrt{2/\pi}$ and variance $v_0=1-2/\pi$.

With $\xi_i=R_i-m_0$, $D_f=\diag(f)$, and $a=\rho m_0$,
\[
 \sigma\rho HX+sY
 =\underbrace{\sigma aHf}_{\text{fixed given }f}
  +\sigma\rho HD_f\xi+sY.
\]
Replace each $\xi_i$ by an independent centered Gaussian of variance $v_0$, also independent of $f,Y$. These replacements match the first two moments of $\xi_i$. Conditional on $f$, the transformed replacement vector is jointly Gaussian with covariance $v_0HD_fD_f^TH^T=v_0I$. Its coordinates are independent, and its law does not depend on $f$. Adding the independent noise $sY$ gives the \emph{conditional product distribution}
\[
 \Pcal_{n,\sigma}:\quad f\text{ uniform},\qquad
       g=\sgn(\sigma aHf+bZ),\qquad
 b^2=\rho^2v_0+s^2=1-a^2,
\]
where $Z\sim\Normal(0,I_D)$ is independent of $f$. Section~\ref{sec:shared-analysis} bounds the change in test expectations under this replacement for the tests in Sections~\ref{sec:ip} and~\ref{sec:ph}.

\paragraph{Uniform positions conditional on a count.}
Conditional independence makes the positions within each bin easy to sample. Let $\varphi(z)=(2\pi)^{-1/2}e^{-z^2/2}$ and $\Phi(z)=\int_{-\infty}^z\varphi(t)\,dt$ be the standard normal density and distribution function. Under $\Pcal_{n,+}$, every coordinate in a weight bin $I_w(f)$ has the same plus probability
\[
 \Pp(g_i=+1\mid f)=\Phi\!\left(\frac{a w}{b\sqrt D}\right)
 \qquad (i\in I_w(f)).
\]
The sign vectors with exactly $k$ plus entries form the $k$-\emph{slice}. Conditional on its plus count, each bin is uniform on the corresponding slice. Repair changes the count law while preserving this conditional placement law.

\paragraph{Calibrating the energy target.}
We choose $\rho$ to bring the reference energy near $S_*$ and leave a small discrepancy to repair. For $\tau>0$, define
\begin{equation}\label{eq:gaussian-link}
 h_\tau(z)=2\Phi(z/\tau)-1.
\end{equation}
This is the mean of $\sgn(z+\tau Z)$ for a standard normal $Z$. In particular, $\E[g_i\mid f]=\sigma h_b(a(Hf)_i)$ under $\Pcal_{n,\sigma}$. Since $u_i=\sqrt D(Hf)_i$, the normalized positive conditional mean energy is
\[
 \frac{\E_{\Pcal_{n,+}}[S(f,g)\mid f]}{D^{3/2}}
 =\frac1D\sum_i(Hf)_i h_b(a(Hf)_i).
\]
Each $(Hf)_i$ has the distribution of a normalized sum of $D$ independent fair signs. We therefore choose the parameters by requiring the corresponding Gaussian expectation $\E[Zh_b(aZ)]$ to equal the target $1/2$. The approximation error for Boolean $f$ will be bounded in Lemma~\ref{lem:mean}.

For independent standard normals $Z,Z'$, put $V=aZ+bZ'$ and $U=bZ-aZ'$. This orthogonal change of variables makes $V,U$ independent standard normals, with $Z=aV+bU$. Thus $\E[Z\mid V]=aV$. Conditioning first on $Z$, and then on $V$, gives
\begin{equation}\label{eq:calibration}
 \E[Zh_b(aZ)]=\E[Z\sgn V]
          =a\E[V\sgn V]=a\E|V|=\rho(2/\pi).
\end{equation}
The target therefore determines $\rho=\pi/4$. From now on we fix
\begin{equation}\label{eq:constants}
 \begin{aligned}
 m_0&=\sqrt{2/\pi},& v_0&=1-2/\pi,& \rho&=\pi/4,\\
 s&=\sqrt{1-\rho^2},& a&=\rho m_0,& b&=\sqrt{1-a^2}.
 \end{aligned}
\end{equation}
Here $a$ is the conditional signal coefficient, $s$ is the retained independent noise scale, and $b$ is the total noise scale after replacement. In particular $a^2=\pi/8$ and $b,s>0$. Unless stated otherwise, implicit constants may depend on these fixed distribution parameters, but not on the test, side-oracle assignment, or length.

\subsection{Weight bins and their sizes}\label{subsec:weight-bins}

Increasing the count in a weight-$w$ bin by one changes $S(f,g)$ by $2w$. We therefore need weights that permit an exact integer correction, with enough coordinates in each bin to accommodate it. We choose small positive weights in an arithmetic progression for this purpose.

The selected weights depend on $f$ because all coefficients $u_i$ share a residue modulo four. If $f$ has $k$ negative entries, then
\begin{equation}\label{eq:residue}
 u_i\equiv-2k\pmod4\qquad\text{for every }i.
\end{equation}
Indeed, for $n\ge2$, every row of $W$ has an even number of negative entries, namely $D/2$ for a nonzero row index and none for index zero. Multiplying $f$ coordinatewise by a row therefore flips an even number of signs, so the resulting number $k_i$ of negative entries satisfies $k_i\equiv k\pmod2$. Since $4\mid D$, we obtain $u_i=D-2k_i\equiv-2k\pmod4$. Thus nonempty bins require weights congruent to $0$ when $k$ is even, or $2$ when $k$ is odd. Weights in the other class give empty bins.

Let $r\in\{0,2\}$ be the common residue. Set $M=\lfloor D^{1/12}\rfloor$ and $J=M^4$, so that $D=\Theta(M^{12})$. We take the smallest positive weight in the appropriate residue class and then increase by four, setting
\begin{equation}\label{eq:bins}
 w_0=\begin{cases}4,&r=0,\\2,&r=2,\end{cases}
 \qquad w_j=w_0+4j,\qquad
 I_j=\{i:u_i=w_j\},\quad N_j=|I_j|,\quad 0\le j<J.
\end{equation}
The first weight $w_0$ divides every $u_i$ and will allow us to remove the final remainder exactly. The progression supplies many weights in the required residue class. Moreover,
\begin{equation}\label{eq:bin-weight-sum}
 \max_jw_j=O(M^4)=o(\sqrt D),\qquad
 W_\Sigma:=\sum_jw_j=\Theta(J^2)=\Theta(M^8).
\end{equation}
The next lemma ensures sufficient capacity in every selected bin simultaneously, despite the dependence among the Fourier coefficients.

\begin{lemma}[Bin sizes]\label{lem:bins}
There are fixed constants $c,C>0$ such that
\[
 \Pp_f\bigl[c\sqrt D\le N_j\le C\sqrt D
                   \text{ for every }0\le j<J\bigr]
 =1-O(M^{-2}).
\]
\end{lemma}

\begin{proof}
Condition on $r$, fixing all selected weights, and let $p_m(w)=2^{-m}\binom{m}{(m+w)/2}$ be the probability that $m$ independent fair signs sum to $w$.

Fix $w=w_j$ and distinct rows $i\ne i'$. Orthogonality implies that the rows agree on $D/2$ coordinates and disagree on $D/2$. Let $A,B$ be the contributions to $u_i$ from these two sets. Before conditioning, they are independent sums of $D/2$ fair signs, with $u_i=A+B$ and $u_{i'}=A-B$. Thus $u_i=u_{i'}=w$ exactly when $A=w$ and $B=0$. Each membership event below forces the selected residue $r$, which has probability $1/2$. Hence
\[
\begin{aligned}
 q_w:=\Pp(u_i=w\mid r)
      =2p_D(w),\quad t_w:=\Pp(u_i=u_{i'}=w\mid r)
      =2p_{D/2}(w)p_{D/2}(0).
\end{aligned}
\]

All selected weights are even and satisfy $w_j\le4M^4=O(D^{1/3})$. Uniformly in this range, Stirling's formula gives, for $m\in\{D,D/2\}$,
\[
 p_m(w)=\sqrt{\frac{2}{\pi m}}e^{-w^2/(2m)}
 \left[1+O\!\left(\frac1m+\frac{w^2}{m^2}+\frac{w^4}{m^3}\right)\right].
\]
The bracket is $1+O(D^{-1})$, so the additive error is $O(D^{-3/2})$. Consequently,
\[
\begin{aligned}
 q_w=\Theta(D^{-1/2}),\qquad
 q_w^2,t_w=\frac8{\pi D}e^{-w^2/D}+O(D^{-3/2}).
\end{aligned}
\]
The common leading term gives the covariance cancellation
\[
 t_w-q_w^2=O(D^{-3/2}).
\]
Thus the sum of the off-diagonal covariance terms is $O(\sqrt D)$. Writing $N_j$ as a sum of membership indicators gives
\[
\begin{aligned}
 \E[N_j\mid r]
 &=Dq_w=\Theta(\sqrt D),\\
 \operatorname{Var}(N_j\mid r)
 &=Dq_w(1-q_w)+D(D-1)(t_w-q_w^2)=O(\sqrt D).
\end{aligned}
\]
Choose fixed $c,C>0$ so that every conditional mean lies inside $[c\sqrt D,C\sqrt D]$ at distance $\Omega(\sqrt D)$ from both endpoints. Chebyshev's inequality and a union bound over the $J$ bins give failure probability $O(J/\sqrt D)=O(M^{-2})$, uniformly in $r$.
\end{proof}

\paragraph{Choice of scales.}
Increasing $J$ enlarges the available repair weight $W_\Sigma=\Theta(J^2)$, but also increases the simultaneous bin-size failure probability and the binomial fluctuations. The choices $M=\lfloor D^{1/12}\rfloor$ and $J=M^4$ give
\[
 \frac{J}{\sqrt D}=O(M^{-2}),\qquad
 \frac{M^{13}}{W_\Sigma\sqrt D}=O(M^{-1}),\qquad
 \sqrt{\frac{J}{\sqrt D}}=O(M^{-1}).
\]
These control, respectively, simultaneous bin-size failure, the fractional common count correction at the good-data residual cutoff $M^{13}$, and the binomial fluctuation term in Section~\ref{sec:coupling}. The final integer adjustments also contribute $O(J/\sqrt D)$ (Lemma~\ref{lem:counts}). Thus bin-size failure is one power of $M$ below the leading repair error, and $M^{-1}=\Theta(2^{-n/12})$ gives the decay rate in Theorem~\ref{thm:finite-common}. Table~\ref{tab:scales} records the resulting bounds.

\subsection{Concentration of the conditional mean energy}\label{subsec:energy-calibration}

Under $\Pcal_{n,+}$, the energy varies with both $f$ and the signs of $g$ given $f$. We separate these two sources of discrepancy from $S_*=D^{3/2}/2$ by writing
\[
 S(f,g)-S_*
 =\underbrace{\bigl(\E[S(f,g)\mid f]-S_*\bigr)}_{\text{conditional-mean discrepancy}}
  +\underbrace{\bigl(S(f,g)-\E[S(f,g)\mid f]\bigr)}_{\text{fluctuation in }g}.
\]
This subsection bounds the first term in mean square. Section~\ref{subsec:count-repair} combines it with fluctuations of the outside signs and rounding errors.

Since $u_i=\sqrt D(Hf)_i$ and $\E[g_i\mid f]=h_b(a(Hf)_i)$, we have
\[
 \E_{\Pcal_{n,+}}[S(f,g)\mid f]
 =\sum_i u_i\E[g_i\mid f]
 =\sqrt D\sum_i(Hf)_i h_b(a(Hf)_i).
\]
Define the scalar function and its average by
\[
 \psi(z)=zh_b(az),\qquad
 \mathcal A(f)=\frac1D\sum_i\psi((Hf)_i).
\]
Thus $\mathcal A(f)$ is the normalized conditional mean energy, with
\begin{equation}\label{eq:conditional-energy}
 \E_{\Pcal_{n,+}}[S(f,g)\mid f]=D^{3/2}\mathcal A(f).
\end{equation}
It therefore suffices to show $\E_f(\mathcal A(f)-1/2)^2=O(D^{-1})$, giving an $O(D^2)$ mean-square bound for the conditional-mean energy discrepancy. We compare the mean of $\mathcal A$ with the Gaussian expectation calibrated earlier by moment replacement, and bound its variance using orthogonality and the Boolean-cube Poincar\'e inequality.

We record the following two-moment Taylor identity for the magnitude replacement in Section~\ref{sec:shared-analysis}, following the Lindeberg method~\cite{Chatterjee}. The immediate mean estimate uses its fourth-order analogue because fair signs and standard normals also have matching third moments.

\begin{lemma}[Second-order Taylor replacement]\label{lem:taylor}
Let $F:\R\to\R$ be $C^3$, and let $U,V$ have the same mean and second moment. Put
\[
 R(z)=\frac{z^3}{2}\int_0^1(1-t)^2F'''(tz)\,dt.
\]
Whenever the expectations are finite,
\[
 \E[F(U)-F(V)]=\E[R(U)-R(V)].
\]
If $|F'''|\le B$, the absolute difference is at most $B(\E|U|^3+\E|V|^3)/6$.
\end{lemma}
\begin{proof}
Taylor's formula with integral remainder gives the identity because the terms of degrees one and two have matching expectations. The bound follows by integrating $(1-t)^2$.
\end{proof}
If the test also depends on random inputs independent of $(U,V)$, apply the identity conditionally on those inputs and then average.

\begin{lemma}[Poincar\'e inequality~{\cite[p.~52]{ODonnell}}]
\label{lem:cube-variance}
For uniform $f\in\{\pm1\}^m$ and $A:\{\pm1\}^m\to\R$,
\[
 \operatorname{Var}(A(f))
 \le \frac14\sum_{i=1}^m
 \E\bigl(A(f^{(i)})-A(f)\bigr)^2,
\]
where $f^{(i)}$ is $f$ with its $i$-th sign flipped.
\end{lemma}
\begin{lemma}[Mean-square energy calibration]\label{lem:mean}
For uniform $f$,
\[
 \E\mathcal A(f)=1/2+O(D^{-1}),\qquad
 \operatorname{Var}\mathcal A(f)=O(D^{-1}),
\]
and consequently
\begin{equation}\label{eq:energy-mse}
 \E\bigl(\mathcal A(f)-1/2\bigr)^2=O(D^{-1}).
\end{equation}
\end{lemma}
\begin{proof}
The derivatives $\psi^{(k)}$, $1\le k\le4$, are bounded. Indeed, with $c=a/b$, we have $\psi'(z)=2\Phi(cz)-1+2cz\varphi(cz)$, and every derivative of order at least two is a polynomial in $z$ times $\varphi(cz)$. Gaussian decay bounds each such product.

To estimate the mean, observe that multiplying independent fair signs by a fixed row of $W$ preserves their distribution. Each $(Hf)_i$ therefore has the distribution of $D^{-1/2}\sum_{j=1}^D\epsilon_j$, where the $\epsilon_j$ are independent fair signs. By linearity of expectation,
\[
 \E\mathcal A(f)
 =\E\psi\!\left(D^{-1/2}\sum_{j=1}^D\epsilon_j\right).
\]

We replace the $\epsilon_j$ one at a time by independent standard normals. At each step, condition on the normalized sum $T$ of the other $D-1$ terms and put $F(t)=\psi(T+D^{-1/2}t)$. A fair sign $\epsilon$ and a standard normal $Z$ have the same first three moments $0,1,0$. We may therefore expand through degree three, one order beyond Lemma~\ref{lem:taylor}. Since $\|F^{(4)}\|_\infty\le D^{-2}\|\psi^{(4)}\|_\infty$, the fourth-order remainder gives, uniformly in $T$,
\[
 \bigl|\E[F(\epsilon)-F(Z)\mid T]\bigr|
 \le\frac{\|\psi^{(4)}\|_\infty}{24D^2}
       (\E|\epsilon|^4+\E|Z|^4)
 =\frac{\|\psi^{(4)}\|_\infty}{6D^2}.
\]
The last equality uses $\E|\epsilon|^4=1$ and $\E|Z|^4=3$. After $D$ replacements the normalized sum is standard normal, so summing the errors and using \eqref{eq:calibration} yields
\[
 \E\mathcal A(f)=\E\psi(Z)+O(D^{-1})=1/2+O(D^{-1}).
\]

For the variance, the dependence among the coordinates of $Hf$ matters. We instead control changes in the independent input signs $f_i$. Extend $\mathcal A$ by the same formula to $[-1,1]^D$ so that we can differentiate along sign-flip segments. The chain rule, with $\psi'$ evaluated coordinatewise, gives
\[
 \partial_i\mathcal A(x)
 =\frac1D\sum_jH_{ji}\psi'((Hx)_j),\qquad
 \nabla\mathcal A(x)=\frac1D H^T\psi'(Hx).
\]
Orthogonality of $H$ then gives
\[
 \|\nabla\mathcal A(x)\|_2^2
 =\frac1{D^2}\sum_j\psi'((Hx)_j)^2
 \le\frac{\|\psi'\|_\infty^2}{D}.
\]
To bound sign flips, we also control the Taylor remainder. Since $H_{ji}^2=D^{-1}$,
\[
 \partial_{ii}\mathcal A(x)
 =\frac1D\sum_jH_{ji}^2\psi''((Hx)_j)
 =\frac1{D^2}\sum_j\psi''((Hx)_j),\qquad
 |\partial_{ii}\mathcal A(x)|\le\frac{\|\psi''\|_\infty}{D}.
\]
Since the flip $f^{(i)}=f-2f_i e_i$ moves distance two in coordinate $i$, Taylor's formula gives
\[
 \mathcal A(f^{(i)})-\mathcal A(f)
 =-2f_i\partial_i\mathcal A(f)+R_i,\qquad
 |R_i|\le\frac{2\|\psi''\|_\infty}{D}.
\]
Applying Lemma~\ref{lem:cube-variance} and $(x+y)^2\le2x^2+2y^2$, we obtain
\[
 \begin{aligned}
 \operatorname{Var}\mathcal A(f)
 &\le2\E\|\nabla\mathcal A(f)\|_2^2
       +\frac12\sum_i\E R_i^2\\
 &\le\frac{2(\|\psi'\|_\infty^2+\|\psi''\|_\infty^2)}{D}.
 \end{aligned}
\]
Combining the variance bound with the squared bias proves \eqref{eq:energy-mse}, since
\[
 \E(\mathcal A(f)-1/2)^2
 =\operatorname{Var}\mathcal A(f)
   +(\E\mathcal A(f)-1/2)^2
 =O(D^{-1})+O(D^{-2})=O(D^{-1}).\qedhere
\]
\end{proof}

Returning to the unnormalized energy using \eqref{eq:conditional-energy}, we have
\[
 \E_f\bigl[(\E_{\Pcal_{n,+}}[S(f,g)\mid f]-S_*)^2\bigr]
 =D^3\E_f(\mathcal A(f)-1/2)^2=O(D^2).
\]
Thus the conditional-mean discrepancy has root-mean-square size $O(D)$.

\subsection{Repairing the bin counts}\label{subsec:count-repair}

Draw $f$ uniformly and put $I=\bigcup_jI_j$. Draw only the signs $(g_i)_{i\notin I}$ from $\Pcal_{n,+}$ conditional on $f$, and record the \emph{outer data} and outside energy
\[
 \omega=\bigl(f,(g_i)_{i\notin I}\bigr),\qquad
 S_{\rm out}=\sum_{i\notin I}u_i g_i.
\]
We choose all counts deterministically from $\omega$. Under $\Pcal_{n,+}$, the signs on $I$ remain independent conditional on this outer data, which includes $f$. Within bin $I_j$, their common plus probability is
\begin{equation}\label{eq:pj}
 p_j=\Phi\left(\frac{a w_j}{b\sqrt D}\right)
     =1/2+O(M^{-2}),
\end{equation}
since $w_j/\sqrt D=O(M^{-2})$. Thus eventually $p_j\in[2\alpha,1-2\alpha]$ for a fixed $\alpha>0$. This margin from zero and one will allow count changes in either direction.

\paragraph{Baseline counts and residual energy.}
The conditional expected plus count in bin $I_j$ is $N_jp_j$. Fix a nearest-integer rounding rule, including its treatment of ties, and put $K_j^0=\operatorname{round}(N_jp_j)$. These rounded means satisfy $|K_j^0-N_jp_j|\le1/2$. The baseline energy $S_0$ and \emph{residual} $\Delta$ are
\begin{equation}\label{eq:residual}
 S_0=S_{\rm out}+\sum_j w_j(2K_j^0-N_j),\qquad
 \Delta=(S_*-S_0)/2.
\end{equation}
Each count increment changes the energy by twice its weight, so the corrected counts $K_j=K_j^0+d_j$ achieve the target precisely when
\[
 \sum_jw_jd_j=\Delta.
\]
\begin{lemma}[Residual energy before repair]\label{lem:residual}
When the outer data $\omega$ have the law induced by $\Pcal_{n,+}$,
\begin{equation}\label{eq:residual-mse}
 \E\Delta^2=O(D^2)=O(M^{24}).
\end{equation}
\end{lemma}
\begin{proof}
Put $Z_{\rm out}=S_{\rm out}-\E[S_{\rm out}\mid f]$ and $B(f)=2\sum_jw_j(K_j^0-N_jp_j)$. The conditional expected contribution of bin $I_j$ is $w_j(2N_jp_j-N_j)$. Consequently, \eqref{eq:conditional-energy} gives
\[
 S_0-S_*=D^{3/2}(\mathcal A(f)-1/2)+Z_{\rm out}+B(f).
\]
The three terms are the conditional-mean discrepancy, the fluctuation in the outside signs, and the rounding error. Lemma~\ref{lem:mean} gives an $O(D^2)$ mean-square bound for the first. For the second, conditional independence gives
\[
 \E[Z_{\rm out}^2\mid f]
 =\sum_{i\notin I}u_i^2\operatorname{Var}(g_i\mid f)
 \le\sum_i u_i^2
 =D\|f\|_2^2=D^2,
\]
where we used $\operatorname{Var}(g_i\mid f)\le1$ and $W^TW=D I_D$. Finally, nearest-integer rounding gives
\[
 |B(f)|\le2\sum_jw_j\,|K_j^0-N_jp_j|
          \le W_\Sigma=O(M^8).
\]
Using $(x+y+z)^2\le3(x^2+y^2+z^2)$ and $\Delta=-(S_0-S_*)/2$ proves the claim, since $M^{16}=O(D^2)$.
\end{proof}

Let $\Good$ be the event, determined by the outer data, that all bin-size bounds hold and $|\Delta|\le M^{13}$. This cutoff is a factor $M$ larger than the residual's root-mean-square scale $O(M^{12})$. Markov's inequality gives $\Pp(|\Delta|>M^{13})\le\E\Delta^2/M^{26}=O(M^{-2})$. Together with Lemma~\ref{lem:bins}, this yields
\begin{equation}\label{eq:outer-good}
 \Pp(\Good^c)=O(M^{-2}).
\end{equation}

\paragraph{Integer balancing.}
A common shift of all counts uses the large sum $W_\Sigma$ to remove most of the residual. We then adjust the largest-weight bin to reduce the remainder below one weight, and use the smallest-weight bin to remove that remainder exactly. The proof first checks the necessary divisibility, then shows that these corrections remain small relative to every bin's size.

\begin{lemma}[The repaired counts]\label{lem:counts}\label{lem:integer}
For every good outer datum there are deterministically chosen integers $K_j$ satisfying
\begin{align}
 \sum_jw_j(2K_j-N_j)&=S_*-S_{\rm out},\label{eq:exact-counts}\\
 |K_j-N_jp_j|&\le CM^5.\label{eq:count-error}
\end{align}
For large $D$, $K_j/N_j\in[\alpha,1-\alpha]$.
\end{lemma}
\begin{proof}
We first check that the residual has the required divisibility. Since every weight is a multiple of $w_0$, an integer solution requires $\Delta\in w_0\mathbb Z$. Every $u_i$ is also a multiple of $w_0$, so $u_i g_i\equiv u_i\pmod{2w_0}$ for either sign $g_i$. Moreover,
\[
 \sum_i u_i=\sum_x f(x)\sum_i(-1)^{i\cdot x}=Df(0),
\]
since the inner sum is $D$ at $x=0$ and zero otherwise. Thus every completion has energy congruent to $Df(0)$ modulo $2w_0$. The baseline counts lie in $[0,N_j]$, so $S_0$ is the energy of such a completion. At eligible lengths, $n$ is a multiple of four and $n\ge4$, so $D=2^n$ and $S_*=2^{3n/2-1}$ are both divisible by $2w_0\in\{4,8\}$. Hence $S_*-S_0$ is divisible by $2w_0$, and dividing by two proves $\Delta\in w_0\mathbb Z$.

We now choose the counts. Using the fixed rounding rule, set
\[
 \begin{aligned}
 t&=\operatorname{round}(\Delta/W_\Sigma),
 &\Delta_1&=\Delta-tW_\Sigma,\\
 a_1&=\operatorname{round}(\Delta_1/w_{J-1}),
 &c_1&=(\Delta_1-a_1w_{J-1})/w_0.
 \end{aligned}
\]
Since $W_\Sigma$ and $w_{J-1}$ are multiples of $w_0$, the divisibility just proved makes $c_1$ an integer. Set
\[
 d_j=t+a_1\one_{\{j=J-1\}}+c_1\one_{\{j=0\}},
 \qquad K_j=K_j^0+d_j.
\]
Then
\[
 \sum_jw_jd_j=tW_\Sigma+a_1w_{J-1}+c_1w_0=\Delta,
\]
which gives \eqref{eq:exact-counts}.

It remains to check that the corrections fit within the bins. Nearest-integer rounding leaves $|\Delta_1|\le W_\Sigma/2$ and $|\Delta_1-a_1w_{J-1}|\le w_{J-1}/2$. On $\Good$, therefore,
\[
 \begin{aligned}
 |t|&\le\frac{|\Delta|}{W_\Sigma}+\frac12=O(M^5),\\
 |a_1|&\le\frac{W_\Sigma}{2w_{J-1}}+\frac12=O(J),\\
 |c_1|&\le\frac{w_{J-1}}{2w_0}=O(J).
 \end{aligned}
\]
Since $J=M^4$, each $|d_j|=O(M^5)$, and $|K_j-N_jp_j|\le|d_j|+1/2$ proves \eqref{eq:count-error}. Finally, $N_j=\Theta(M^6)$ on $\Good$, so
\[
 \left|\frac{K_j}{N_j}-p_j\right|=O(M^{-1}).
\]
The margin $p_j\in[2\alpha,1-2\alpha]$ therefore ensures $K_j/N_j\in[\alpha,1-\alpha]$ for large $D$.
\end{proof}

\begin{table}[!htb]
\centering\normalsize
\begin{tabular}{@{}p{0.27\textwidth}p{0.20\textwidth}p{0.45\textwidth}@{}}
\toprule
Quantity & Scale & Role in the construction\\
\midrule
Ambient size $D$ & $\Theta(M^{12})$ & Relates the scales to input length\\[0.4em]
Number of bins $J$ & $M^4$ & Many adjustable count coordinates\\[0.4em]
Largest weight $w_{J-1}$ & $\Theta(M^4)$ & Bounds the final integer corrections\\[0.4em]
Bin size $N_j$ & $\Theta(M^6)$ & Capacity and random placement\\[0.4em]
Weight sum $W_\Sigma$ & $\Theta(M^8)$ & Converts residual to a common shift\\[0.4em]
Allowed residual $|\Delta|$ & $\le M^{13}$ & Cutoff exceeded with probability $O(M^{-2})$\\[0.4em]
Count correction & $O(M^5)$ & $M^{13}/M^8+O(M^4)$\\[0.4em]
Fractional correction & $O(M^{-1})$ & Count correction divided by bin size\\
\bottomrule
\end{tabular}
\caption{Scales on $\Good$ for count repair. Bin-size and count-correction bounds apply to every selected bin.}
\label{tab:scales}
\end{table}
\subsection{\texorpdfstring{Sampling the exact distributions $\Ecal_{n,\sigma}$}{Sampling the exact distributions E(n,sigma)}}\label{subsec:exact-sampling}

Choose $n_0\in\Nelig$ so that the preceding feasibility bounds hold for every eligible $n\ge n_0$. For eligible $n<n_0$, let $\Ecal_{n,+}$ be the point mass at the default pair \eqref{eq:filler}; these finitely many lengths do not affect the asymptotic bounds. For eligible $n\ge n_0$, use the following procedure.
\begin{enumerate}[label=(\arabic*)]
\item Draw $f$ uniformly, determine the selected weight bins $I_j$, and put $I=\bigcup_j I_j$.
\item Draw the signs $g_i$, $i\notin I$, from $\Pcal_{n,+}$ conditional on $f$. This gives the outer data $\omega=(f,(g_i)_{i\notin I})$. Leave all signs on $I$ unexposed.
\item Compute the baseline counts and residual from \eqref{eq:residual}, and test the bin-size bounds and $|\Delta|\le M^{13}$ defining $\Good$.
\item On $\Good$, choose the deterministic counts $K_j$ from Lemma~\ref{lem:counts}. Independently in each bin, choose a uniform $K_j$-subset of $I_j$ for its plus positions, set the remaining signs to minus, and output $(f,g)$.
\item On $\Good^c$, discard both $f$ and the sampled outside signs, and output the default pair from \eqref{eq:filler}.
\end{enumerate}
Here $\Good$ is an event of the initially sampled outer data.

To sample $\Ecal_{n,-}$, draw $(f,g)$ from $\Ecal_{n,+}$ and output $(f,-g)$. The two distributions share the same marginal law for $f$, which may differ from uniform because of the default replacement. Since the output $f$ equals the initially sampled uniform vector outside $\Good^c$, this coupling bounds its total-variation distance from uniform by $\Pp(\Good^c)=O(M^{-2})$.

\begin{proposition}[Properties of $\Ecal_{n,\sigma}$]\label{prop:law-summary}
For all sufficiently large $n\in\Nelig$, the distributions $\Ecal_{n,\sigma}$, $\sigma\in\{+1,-1\}$, satisfy $\operatorname{supp}(\Ecal_{n,\sigma})\subseteq\mathcal L_{n,\sigma}$. In the positive construction, $\Pp(\Good^c)=O(M^{-2})$, and on $\Good$,
\[
 J=M^4,\quad N_j=\Theta(M^6),\quad W_\Sigma=\Theta(M^8),\quad
 p_j\in[2\alpha,1-2\alpha],\quad |K_j/N_j-p_j|\le C/M.
\]
Conditional on the initially sampled outer data $\omega\in\Good$, the signs on $I$ under $\Pcal_{n,+}$ are independent; under $\Ecal_{n,+}$, the sign vectors in the bins are independent and uniform on their prescribed-count slices. These assertions concern the conditional sampling rules before any default replacement.
\end{proposition}
\begin{proof}
The quantitative bounds follow from Lemmas~\ref{lem:bins}, \ref{lem:residual}, and~\ref{lem:counts}, together with \eqref{eq:pj} and~\eqref{eq:bin-weight-sum}. On $\Good$, every choice of positions with the prescribed counts satisfies
\[
 S(f,g)=S_{\rm out}+\sum_j w_j(2K_j-N_j)=S_*
\]
by \eqref{eq:exact-counts}, hence $\forr(f,g)=1/2$. On $\Good^c$, the default input belongs to $\mathcal L_{n,+}$ by \eqref{eq:filler}. Finally, $\forr(f,-g)=-\forr(f,g)$, giving the negative support. The conditional independence and uniform-slice assertions follow from the sampling rule.
\end{proof}

\section{\texorpdfstring{Comparing $\Ecal_{n,\sigma}$ with $\Pcal_{n,\sigma}$}{Comparing E(n,sigma) with P(n,sigma)}}\label{sec:coupling}

We bound the effect of count repair on two families of classical tests: convex tests arising from interactive verifiers in Section~\ref{sec:ip}, and Boolean tests whose probability of remaining nonconstant under a fair random restriction of density $t$ is at most $t\Lambda$, as in Section~\ref{sec:ph}.

We first take $\sigma=+1$ and fix good outer data $\omega=(f,(g_i)_{i\notin I})$ from Proposition~\ref{prop:law-summary}, where $I=\bigcup_j I_j$. The bins, plus probabilities, and prescribed counts are fixed, with
\[
 J=M^4,\qquad N_j=\Theta(M^6),\qquad
 p_j,K_j/N_j\in[\alpha,1-\alpha],\qquad
 |K_j/N_j-p_j|\le C_0/M,
\]
where $\alpha>0$ is fixed and $C_0$ is a sufficiently large absolute constant. Under $\Pcal_{n,+}$, the counts $\ell_j\sim\mathrm{Bin}(N_j,p_j)$ are independent, while under $\Ecal_{n,+}$, each count is $K_j$. Conditional on the counts, both distributions choose the plus positions uniformly and independently across bins.

Small fractional count discrepancy does not imply total-variation closeness. For example, the event ``the plus count equals $K$'' has probability one on the $K$-slice, but probability at most $O(N^{-1/2})$ for $\mathrm{Bin}(N,p)$ when $p\in[\alpha,1-\alpha]$. Thus even a slice centered at the product mean can be far from the product law in total variation. We instead use $|K_j/N_j-p_j|=O(M^{-1})$ to compare expectations for the specified test families.

With the outer data fixed, we regard each test as a function of the remaining signs and keep its notation unchanged. Convexity and the Lipschitz bound survive these assignments. We require the same stability of the Boolean assumption below.

For a real function on a finite set $S$, write $\osc_S F=\max_S F-\min_S F$. We say that $F$ is $L_1$-Lipschitz in the infinity norm if
\begin{equation}\label{eq:lip}
 |F(x)-F(y)|\le L_1\norm{x-y}_\infty.
\end{equation}
For Boolean tests, we use the restriction notation of Section~\ref{subsec:test-models}. Here $\Res_t$ leaves each coordinate free with probability $t$ and otherwise fixes a fair sign. Only the order-one case of Definition~\ref{def:criticality} is needed here. We assume that for every deterministic restriction $\tau$ and every $t\in[0,1]$,
\begin{equation}\label{eq:first-criticality}
 \Pp_{\eta\sim\Res_t}(C|_{\tau,\eta}\text{ is nonconstant})\le t\Lambda.
\end{equation}
The uniformity in $\tau$ allows us to fix the outer data, and later make further assignments, without changing $\Lambda$.

\begin{proposition}[Stability under count repair]\label{prop:repair}
For all sufficiently large $n\in\Nelig$ and each $\sigma\in\{+1,-1\}$, every convex function $F:[-1,1]^{2D}\to\R$ that is $L_1$-Lipschitz in infinity norm satisfies
\begin{equation}\label{eq:convex-repair}
 |\E_{\Ecal_{n,\sigma}}F-\E_{\Pcal_{n,\sigma}}F|
 \le C_0\left(\frac{L_1}{M}
       +\frac{\osc_{\{\pm1\}^{2D}}F}{M^2}\right).
\end{equation}
Every Boolean function $C:\{\pm1\}^{2D}\to\{0,1\}$ satisfying \eqref{eq:first-criticality} obeys
\begin{equation}\label{eq:boolean-repair}
 |\E_{\Ecal_{n,\sigma}}C-\E_{\Pcal_{n,\sigma}}C|
 \le C_0\left(\frac{\Lambda}{M}+\frac1{M^2}\right).
\end{equation}
\end{proposition}

The $M^{-1}$ terms measure repair on good data, while the $M^{-2}$ terms account for the default replacement, since $\Pp(\Good^c)=O(M^{-2})$. The oscillation factor allows arbitrary real-valued convex tests. We prove the conditional bounds for $\sigma=+1$, then average.

\subsection{Convex tests via conditional means}
Let $X$ denote a reference sample and $Y$ its coupled exact-count sample. Even one changed sign gives $\|Y-X\|_\infty=2$, so a direct Lipschitz estimate does not capture the small fraction of changes. Instead we compare through $\E[Y\mid X]$ and $\E[X\mid Y]$. These means lie in the cube and have small expected distances from $X$ and $Y$. Convexity converts these bounds into the two sides of the desired expectation inequality.

We couple the counts within each bin by using initial segments of the same random ordering. The resulting sets of plus positions are uniform and nested. For a bin of size $N$, let $Z(k)$ put plus signs at the first $k$ positions of a uniform permutation. Independently draw $\ell\sim\mathrm{Bin}(N,p)$, and set $X=Z(\ell)$, $Y=Z(K)$. Exactly $|\ell-K|$ coordinates differ, and the ordering spreads these changes uniformly among the eligible positions. This increasing-chain coupling appears in Filmus--Mossel \cite[full version, Definition~4.1 and Section~8]{FM}. We use its permutation description to derive conditional means for a prescribed target count $K$, and its equivalent independent-uniform description to construct the random restriction in the next subsection.

\begin{lemma}[Coupling product signs and a slice]\label{lem:rank}
The vector $X$ has independent signs of plus probability $p$, while $Y$ is uniform on the $K$-slice and independent of $\ell$. Equivalently, take independent uniform $U_i\in[0,1]$, give $X$ plus signs where $U_i\le p$, and give $Y$ plus signs at the $K$ smallest uniforms. For $0<K<N$, the nonzero coordinate magnitudes of $\E[Y\mid X]-X$ are
\begin{equation}\label{eq:forward-drifts}
 \frac{2(\ell-K)_+}{\ell}\text{ on plus positions of }X,\qquad
 \frac{2(K-\ell)_+}{N-\ell}\text{ on minus positions of }X.
\end{equation}
The corresponding magnitudes of $\E[X\mid Y]-Y$ are
\begin{equation}\label{eq:reverse-drifts}
 \frac{2\E[(K-\ell)_+]}{K}\text{ on plus positions of }Y,\qquad
 \frac{2\E[(\ell-K)_+]}{N-K}\text{ on minus positions of }Y.
\end{equation}
\end{lemma}
\begin{proof}
A vector with $a$ plus signs has probability $\Pp(\ell=a)/\binom Na=p^a(1-p)^{N-a}$, proving the product distribution. The permutation is independent of $\ell$, hence so is $Y$. The uniform realization is equivalent because the ordering permutation is independent of the ordered uniform values.

Conditioned on $X$, the differing coordinates form a uniform subset of the eligible plus or minus positions, giving \eqref{eq:forward-drifts}. A zero denominator occurs only when there are no coordinates of the indicated sign, so that expression is unused. Conditioned on $Y$, the count $\ell$ remains binomial, and the reverse subset or superset operation gives \eqref{eq:reverse-drifts} after averaging.
\end{proof}
We couple the bins independently. The fixed outer coordinates contribute zero drift, so the full infinity norm is the maximum of the binwise norms. Conditional Jensen then gives the following comparison.

\begin{lemma}[Convex comparison by conditional means]\label{lem:barycenter}
Let $X,Y$ be coupled random vectors taking values in a cube, and let $F$ be convex and $L_1$-Lipschitz in infinity norm on that cube. Define
\[
 d_\to=\E\norm{\E[Y\mid X]-X}_\infty,\qquad
 d_\leftarrow=\E\norm{\E[X\mid Y]-Y}_\infty.
\]
Then
\begin{equation}\label{eq:barycenter}
 -L_1d_\to\le\E F(Y)-\E F(X)\le L_1d_\leftarrow.
\end{equation}
\end{lemma}
\begin{proof}
Conditional Jensen gives $\E[F(Y)\mid X]\ge F(\E[Y\mid X]) \ge F(X)-L_1\norm{\E[Y\mid X]-X}_\infty$. Averaging proves one inequality, and interchanging $X,Y$ proves the other.
\end{proof}

\begin{proof}[Proof of \eqref{eq:convex-repair}]
With good outer data fixed, $|\ell_j-K_j|/N_j$ is the fraction of differing signs in bin $j$. Let
\[
 \varepsilon=\E\max_j\frac{|\ell_j-K_j|}{N_j}.
\]
In \eqref{eq:forward-drifts}, a changing plus coordinate has $\ell\ge K\ge\alpha N$, and a changing minus coordinate has $N-\ell\ge N-K\ge\alpha N$. The reverse denominators $K,N-K$ in \eqref{eq:reverse-drifts} are also at least $\alpha N$. Thus
\[
 \begin{aligned}
 d_\to,d_\leftarrow&\le\frac{2\varepsilon}{\alpha},\\
 \varepsilon
 &\le\max_j|K_j/N_j-p_j|
       +\left(\sum_j\frac{p_j(1-p_j)}{N_j}\right)^{1/2}\\
 &\le\frac{C_0}{M}+\frac12\left(\sum_jN_j^{-1}\right)^{1/2}\\
 &=\frac{C_0}{M}+O\!\left(\sqrt{M^4/M^6}\right)=O(M^{-1}).
 \end{aligned}
\]
Lemma~\ref{lem:barycenter} now gives the conditional $O(L_1/M)$ bound. We average over the common outer data $\omega$, initially drawn under $\Pcal_{n,+}$, and write $x_*$ for the default input. The exact sample uses $Y$ on $\Good$ and $x_*$ otherwise, whereas the reference sample always uses $X$. Hence
\[
 \begin{aligned}
 \E_{\Ecal_{n,+}}F-\E_{\Pcal_{n,+}}F
 &=\E\!\left[\one_{\Good}
     \bigl(\E[F(Y)\mid\omega]-\E[F(X)\mid\omega]\bigr)\right]\\
 &\quad+\E\!\left[\one_{\Good^c}
     \bigl(F(x_*)-\E[F(X)\mid\omega]\bigr)\right].
 \end{aligned}
\]
Because $\Good$ is determined by $\omega$, its indicators remain inside the expectation over the original outer-data law. No division by $\Pp(\Good)$ is needed. Taking absolute values gives
\begin{equation}\label{eq:conditioning}
 |\E_{\Ecal_{n,+}}F-\E_{\Pcal_{n,+}}F|
 \le\E\!\left[\one_{\Good}|\E[F(Y)\mid\omega]-\E[F(X)\mid\omega]|\right]
       +\Pp(\Good^c)\osc_{\{\pm1\}^{2D}}F.
\end{equation}
The conditional bound and $\Pp(\Good^c)=O(M^{-2})$ prove the claim.
\end{proof}

\subsection{Boolean tests via a common restriction}
If both coupled samples extend a restriction on which the Boolean test is constant, their answers agree. The actual disagreement set need not be independent across coordinates. Using the independent uniforms from Lemma~\ref{lem:rank}, we instead leave fixed windows around the thresholds $p_j$ free. Each restriction coordinate depends only on its own uniform variable, so the resulting restriction is a product restriction. The product sample always extends it; the exact-count sample extends it with probability $1-e^{-\Omega(M^4)}$.

The fixed signs need not be fair, whereas \eqref{eq:first-criticality} assumes fair signs. The next lemma represents this biased restriction as a mixture of fair restrictions. We first sample a preliminary partial assignment, then apply a fair restriction to its remaining coordinates. Write $\DT(C)$ for the minimum worst-case decision-tree depth of $C$. Nonconstancy is exactly the event $\DT(C)\ge1$. Repair only needs this case, but Section~\ref{sec:ph} will use the higher-order statement.

\begin{lemma}[Biased restriction as a mixture]\label{lem:biased-transfer}
Fix a positive integer $r$. Suppose that for every deterministic restriction $\tau$ and $v\in[0,1]$,
\[
 \Pp_{\eta\sim\Res_v}[\DT(C|_{\tau,\eta})\ge r]\le(v\Lambda)^r.
\]
Let $\eta$ be a product restriction that assigns coordinate $i$ to $+1,*,-1$ with probabilities $a_i,s_i,b_i$, respectively, where $*$ leaves the coordinate free. Define
\begin{equation}\label{eq:biased-scale}
 t_*:=\max_i\frac{s_i}{s_i+2\min(a_i,b_i)}
     =\max_i\frac{s_i}{1-|a_i-b_i|},
\end{equation}
with $0/0=0$. Then
\begin{equation}\label{eq:biased-tail}
 \Pp[\DT(C|_\eta)\ge r]\le(t_*\Lambda)^r.
\end{equation}
The assertion also holds after any prior deterministic restriction.
\end{lemma}
\begin{proof}
If $t_*=0$, every coordinate is fixed and the claim is immediate. Otherwise choose independent preliminary assignments with probabilities
\[
 \Pp(\eta_{0,i}=*)=s_i/t_*,\quad
 \Pp(\eta_{0,i}=+1)=a_i-\frac{s_i(1-t_*)}{2t_*},\quad
 \Pp(\eta_{0,i}=-1)=b_i-\frac{s_i(1-t_*)}{2t_*}.
\]
The definition of $t_*$ makes these nonnegative and their sum is one. Applying $\Res_{t_*}$ to the remaining coordinates gives exactly the target probabilities $(a_i,s_i,b_i)$. Conditional on $\eta_0$, the assumed tail bound is at most $(t_*\Lambda)^r$, so averaging proves \eqref{eq:biased-tail}. Any prior deterministic restriction is absorbed into $\eta_0$.
\end{proof}

\begin{proof}[Proof of \eqref{eq:boolean-repair}]
Fix good outer data and choose $\delta=c_{\rm win}/M$, with the fixed constant large enough that $|K_j/N_j-p_j|\le\delta/2$. Eventually $\delta\le\alpha$, so the window lies in $[0,1]$. Its width covers the prescribed-count shift and leaves a count margin of order $N_j\delta=\Theta(M^5)$, larger than the binomial standard deviation $\Theta(M^3)$. Define the restriction by
\[
 \eta_i=
 \begin{cases}
 +1,&U_i<p_j-\delta,\\
 *,&p_j-\delta\le U_i\le p_j+\delta,\\
 -1,&U_i>p_j+\delta,
 \end{cases}
 \qquad i\in I_j.
\]
The product sample $X$ always extends $\eta$, and the exact-count sample $Y$ does so if, in every bin,
\[
 \#\{i:U_i<p_j-\delta\}\le K_j
 \le\#\{i:U_i\le p_j+\delta\}.
\]
The mean of each count is separated from $K_j$ by at least $N_j\delta/2$. For a sum of $N$ independent Bernoulli variables, Hoeffding's inequality bounds either tail at distance $u$ from the mean by $e^{-2u^2/N}$. Since $\min_jN_j\delta^2=\Theta(M^4)$, a union bound yields
\[
 \Pp(Y\text{ does not extend }\eta)\le
 2J e^{-(\min_jN_j)\delta^2/2}=e^{-\Omega(M^4)}.
\]
The restriction probabilities are $(a_i,s_i,b_i)=(p_j-\delta,2\delta,1-p_j-\delta)$. Thus $1-|a_i-b_i|=1-|2p_j-1|=2\min(p_j,1-p_j)$, and $t_*=\max_j\delta/\min(p_j,1-p_j)\le\delta/\alpha$. If $Y$ extends $\eta$ and $C|_\eta$ is constant, the two samples have the same answer. Lemma~\ref{lem:biased-transfer} with $r=1$ therefore gives
\[
 \begin{aligned}
 \Pp(C(X)\ne C(Y))
 &\le\Pp(C|_\eta\text{ is nonconstant})
      +\Pp(Y\text{ does not extend }\eta)\\
 &\le(\delta/\alpha)\Lambda+e^{-\Omega(M^4)}.
 \end{aligned}
\]
Here we apply the restriction bound once to the full remaining test, without conditioning on successful extension. This preserves the product restriction law, and extension failures are covered by the second term. Since $|\E C(X)-\E C(Y)|\le\Pp(C(X)\ne C(Y))$, averaging as in \eqref{eq:conditioning}, with Boolean oscillation at most one, proves the claimed bound.
\end{proof}

Replacing $g$ by $-g$ sends $\Ecal_{n,+}$ to $\Ecal_{n,-}$ by definition and $\Pcal_{n,+}$ to $\Pcal_{n,-}$ by the symmetry $Z\stackrel{d}{=}-Z$. The same statement holds for $\G_{n,\sigma}$ using $Y\stackrel{d}{=}-Y$. Gaussian zeros have probability zero, so the sign convention causes no exception. Composing a test with this coordinate reflection preserves convexity, the Lipschitz bound, and oscillation. It also preserves the restriction bound, since fair fixed signs remain fair and deterministic assignments are carried to deterministic assignments. This proves both repair bounds for $\sigma=-1$.

\section{Analytic tools for the reference comparisons}\label{sec:shared-analysis}

After the count-repair comparison in Section~\ref{sec:coupling}, two steps remain: magnitude replacement from $\Pcal_{n,\sigma}$ to $\G_{n,\sigma}$, and Gaussian comparison from $\G_{n,\sigma}$ to $\Unif_n$. This section develops their shared analytic tools.

Both arguments require smooth functions of the latent real variables. Section~\ref{sec:ip} controls the softmax error and sign-cutoff error for convex tests. For Boolean tests, Section~\ref{sec:ph} averages over independent Gaussian noise already present in the distribution, introducing no error. Those sections supply the derivative bounds for each test family.

\subsection{Derivative norms and Gaussian tails}\label{subsec:comparison-tools}
To exploit the small entries of the Fourier transform matrix,
we bound sums of absolute derivative entries. For vectors,
$\norm{\cdot}_1,\norm{\cdot}_2,\norm{\cdot}_\infty$ are the usual
norms. For matrices and higher tensors, $\norm{\cdot}_1$ denotes
the sum of absolute entries, not an induced operator norm.

For a smooth function $F$ on a neighborhood of $[-1,1]^m$, define
\begin{equation}\label{eq:derivative-budgets}
 \begin{aligned}
 L_r(F)
 :=\sup_{x\in[-1,1]^m}
   \sum_{i_1,\ldots,i_r=1}^m
   \left|\partial_{i_1}\cdots\partial_{i_r}F(x)\right| =\sup_{x\in[-1,1]^m}\norm{\nabla^rF(x)}_1,
 \qquad r=1,2,3.
 \end{aligned}
\end{equation}
The sum runs over all ordered indices, including repetitions.
Writing $D_v^rF(x)=\left.\frac{d^r}{dt^r}F(x+tv)\right|_{t=0}$,
we have, for $x\in[-1,1]^m$ and $r\in\{1,2,3\}$,
\[
 \operatorname{Lip}_\infty(F)\le L_1(F),
 \qquad
 |D_v^rF(x)|\le\norm{\nabla^rF(x)}_1\norm v_\infty^r.
\]
The Lipschitz constant is defined in \eqref{eq:lip}. We write
$L_1,L_2,L_3$ for upper bounds on the quantities $L_r(F)$.

For the $j$th standard basis vector $e_j$, we have
$\norm{He_j}_\infty=D^{-1/2}$. A one-coordinate magnitude
replacement therefore has direction $v_j=\sigma\rho f_jHe_j$
and contributes $\rho^3D^{-3/2}$ to its third-order remainder.
Matching the first two moments cancels the lower Taylor terms
in expectation. Summing over $D$ replacements gives the basic
$D^{-1/2}$ scale; we bound the derivative and moment factors below.
Covariance comparison instead pairs Hessian entries with
covariance changes of size $O(D^{-1/2})$. We therefore need
second- and third-order derivative bounds.

For the latent tests $\Psi,\Psi_f$ introduced below, we use the
following growth convention. \emph{Polynomial growth} of $G$ means
$\norm{G(x)}_1\le C(1+\norm{x}_\infty)^k$ for all $x$,
for some constants $C,k\ge0$, using absolute value for scalars.
The tail estimates below control this growth in expectation.
The constants $c,C$ may change from one occurrence to the next;
subscripts indicate dependence on fixed parameters.

\paragraph{Subgaussian tails and maximum moments.}
A centered real random variable $A$ is $1$-subgaussian if
$\E e^{tA}\le e^{t^2/2}$ for every $t\in\mathbb R$.
Independent centered $1$-subgaussian coordinates $A_i$ satisfy
\[
 \Pp\bigl(|\langle c,A\rangle|>u\bigr)
 \le2\exp\!\left(-\frac{u^2}{2\norm{c}_2^2}\right)
 \qquad(c\ne0,\ u>0).
\]
Since the rows of $H$ have norm one,
$\Pp(\norm{HA}_\infty>u)\le2D e^{-u^2/2}$.

\begin{lemma}[Maximum moments]\label{lem:maxmoment}
Let $V=(V_1,\ldots,V_m)$ satisfy $\Pp(|V_j|>u)\le2e^{-u^2/(2c^2)}$ for some fixed $c>0$, every $j$, and every $u\ge0$. No independence is required. For every fixed $r>0$,
\[
 \E(1+\norm{V}_\infty)^r
 \le C_{r,c}(\log(2m))^{r/2}.
\]
For a suitable fixed $C_c$, $\Pp(\norm V_\infty>C_c\sqrt{\log(2m)})\le m^{-10}$.
\end{lemma}
\begin{proof}
For $T=\norm V_\infty$, the union bound gives
$\Pp(T>u)\le\min\{1,2m e^{-u^2/(2c^2)}\}$.
Integrating this bound after splitting at $2c\sqrt{\log(2m)}$ yields
$\E(1+T)^r\le C_{r,c}(\log(2m))^{r/2}$.
The tail claim follows by taking $u=\sqrt{22}\,c\sqrt{\log(2m)}$.
\end{proof}

\paragraph{A Gaussian tail ratio.}
For $V\sim\Normal(\mu,s^2)$ with $s>0$,
$\min\{\Pp(V>0),\Pp(V<0)\}=\Phi(-|\mu|/s)$.
We use the elementary Mills bound
\begin{equation}\label{eq:mills}
 \Phi(-z)\ge\int_z^{z+(1+z)^{-1}}\varphi(t)\,dt
 \ge\frac{e^{-3/2}\varphi(z)}{1+z}
 \qquad(z\ge0).
\end{equation}
It controls the relative cutoff error in Lemma~\ref{lem:rounding}
and the derivatives of the noisy sign mean in Lemma~\ref{lem:ph-smooth}.

\subsection{Affine covariance interpolation}
The latent Gaussian pair defining $\G_{n,\sigma}$ has covariance $\Sigma_{n,\sigma}$ from \eqref{eq:covariance}. Replacing it by $I_{2D}$ gives independent standard Gaussian coordinates, whose signs have distribution $\Unif_n$. Price's formula compares these endpoints along the straight covariance path, expressing the change in a smooth test expectation through its second derivatives.

\begin{lemma}[Affine Price formula~{\cite[Theorem~1]{Price}}]\label{lem:price}
Let $\Gamma_0,\Gamma_1$ be positive definite covariance matrices. Set $\Gamma_t=(1-t)\Gamma_0+t\Gamma_1$ and $V_t\sim\Normal(0,\Gamma_t)$. If $\Psi$ is $C^2$ and $\Psi$, its gradient, and its Hessian have polynomial growth, then for $0<t<1$,
\begin{equation}\label{eq:price}
 \frac d{dt}\E \Psi(V_t)
 =\frac12\sum_{i,j}(\Gamma_1-\Gamma_0)_{ij}
                          \E\partial_{ij}\Psi(V_t).
\end{equation}
The formula also integrates over $[0,1]$.
\end{lemma}
\begin{proof}
In the upper-triangular covariance coordinates of \cite[Theorem~1]{Price}, differentiating a diagonal entry gives $\frac12\E\partial_{ii}\Psi$, while differentiating an off-diagonal entry gives $\E\partial_{ij}\Psi$. The chain rule along $\Gamma_t$ gives \eqref{eq:price}, since the ordered sum counts each off-diagonal pair twice. The eigenvalue bound
\[
 \lambda_{\min}(\Gamma_t)\ge
 \min\{\lambda_{\min}(\Gamma_0),\lambda_{\min}(\Gamma_1)\}>0
\]
holds uniformly in $t$, and the largest eigenvalue is also uniformly bounded. Thus the Gaussian densities and their covariance derivatives admit integrable bounds after multiplication by the assumed polynomial growth. This justifies continuity at the endpoints and integration over $[0,1]$.
\end{proof}
Taking $\Gamma_0=I_{2D}$ and $\Gamma_1=\Sigma_{n,\sigma}$, the covariance difference vanishes within each block and has cross-block entries of magnitude $\rho/\sqrt D$. Writing the two blocks as $x,y$, the two symmetric cross-block terms cancel the factor $1/2$ in Price's formula, giving
\[
 \frac d{dt}\E\Psi(V_t)
 =\sigma\rho\sum_{i,j}H_{ij}\E\partial_{x_i y_j}\Psi(V_t).
\]
Hence only derivatives involving one coordinate from each block contribute. Integrating \eqref{eq:price} and taking absolute values gives, in particular,
\[
 |\E\Psi(V_1)-\E\Psi(V_0)|
 \le\frac{\rho}{2\sqrt D}\int_0^1
          \E\sum_{i\ne j}|\partial_{ij}\Psi(V_t)|\,dt.
\]
Bounding the \emph{sum} of the relevant Hessian entries therefore retains the small covariance factor without a further factor for the number of pairs.

The Boolean application in Section~\ref{sec:ph} decomposes the Gaussian vector as $V+s_0Y$, with standard normal $Y$ independent of $V$, and averages over $Y$ in the test without approximation error. Subtracting $s_0^2I$ at both endpoints leaves the covariance difference unchanged. The eigenvalue bound $1-\rho>0$ allows a fixed $s_0>0$ for which the remaining covariances stay positive definite. That section specifies $s_0$ and bounds the expected Hessian sum.

The convex application uses Proposition~\ref{prop:convex-gaussian}, which includes its sign-smoothing error. Price's identity itself requires smoothness and growth, not convexity.

\subsection{Replacing centered magnitudes by Gaussians}
Recall from Section~\ref{sec:laws} that $X=D_fR$, with uniform $f$, $D_f=\diag(f)$, and independent half-normal magnitudes $R_i$ independent of $f$. Replacing $\xi_i=R_i-m_0$ by independent $\sqrt{v_0}Z_i$ preserves the conditional signal and the first two noise moments. Before adding $sY$, the endpoints are
\[
 \begin{aligned}
 \text{original:}\quad&\sigma\rho HD_fR
       =\sigma aHf+\sigma\rho HD_f\xi,\\
 \text{replaced:}\quad&\sigma aHf+\rho\sqrt{v_0}\,Z.
 \end{aligned}
\]
Orthogonality makes the replaced Gaussian noise independent of $f$, with covariance $\rho^2v_0I$. Adding $sY$ gives variance $\rho^2v_0+s^2=b^2$ and hence $\Pcal_{n,\sigma}$.

For a Boolean test $C$, the corresponding latent test is
\[
 \Psi_f(\mu)=\E_Y C\bigl(f,\sgn(\mu+sY)\bigr).
\]
Here $\mu$ is the vector before adding noise. Averaging over $f$ and $\mu$ gives the endpoint acceptance expectations. Section~\ref{sec:ph} bounds the derivatives of this exact average; Section~\ref{sec:ip} uses a smooth sign approximation with a controlled error.

We use coordinatewise replacement~\cite[Section~2]{Chatterjee} and the integral Taylor remainder from Lemma~\ref{lem:taylor}. Unlike the fair-sign comparison in Lemma~\ref{lem:mean}, only two moments match, leaving a third-order remainder. We allow a bounded third derivative ($k=0$, for convex tests) or cubic growth in $\norm{\mu}_\infty$ ($k=3$, for Boolean tests). The derivative hypothesis is uniform in $f$; the comparison averages over uniform $f$.

\begin{lemma}[Centered-magnitude replacement]\label{lem:magnitude}
Let $f\in\{\pm1\}^D$ be uniform, let $R=(R_i)_{i=1}^D$ have independent standard half-normal coordinates, and let $Z$ be a standard normal $D$-vector, with $f,R,Z$ mutually independent. Use the constants in \eqref{eq:constants}. For each $f$, let $\Psi_f:\R^D\to\R$ be $C^3$. Suppose, for a fixed $k\in\{0,3\}$, that uniformly in $f,\mu,v$,

\begin{equation}\label{eq:magnitude-hyp}
 |D_v^3\Psi_f(\mu)|
 \le B\norm v_\infty^3(1+\norm\mu_\infty)^k.
\end{equation}

For either $\sigma\in\{+1,-1\}$,
\begin{align}
 &\left|\E\Psi_f(\sigma\rho HD_fR)
       -\E\Psi_f(\sigma aHf+\rho\sqrt{v_0}Z)\right| \le C B D^{-1/2}(\log(2D))^{k/2}.\label{eq:magnitude}
\end{align}

\end{lemma}

\begin{proof}
The derivative hypothesis implies polynomial growth of $\Psi_f$ and its lower derivatives, so all expectations below are finite.
Set $\xi_i=R_i-m_0$ and $\widetilde\xi_i=\sqrt{v_0}Z_i$. They have matching means and variances and bounded sixth moments. Define the hybrids
\[
 \mu^{(r)}=\sigma aHf+\sigma\rho HD_f
   (\widetilde\xi_1,\ldots,\widetilde\xi_r,\xi_{r+1},\ldots,\xi_D)^T,
 \qquad 0\le r\le D.
\]
The endpoint difference telescopes as
\[
 \E\Psi_f(\mu^{(0)})-\E\Psi_f(\mu^{(D)})
 =\sum_{i=1}^D\E\bigl[\Psi_f(\mu^{(i-1)})-\Psi_f(\mu^{(i)})\bigr].
\]
\emph{One-coordinate replacement.} At step $i$, condition on $f$ and all centered variables except $\xi_i,\widetilde\xi_i$. Setting the centered variable at coordinate $i$ to zero gives the base point
\[
 \mu_0=\sigma aHf+\sigma\rho\left(
       \sum_{j<i}f_jH_{\cdot j}\widetilde\xi_j
       +\sum_{j>i}f_jH_{\cdot j}\xi_j\right).
\]
This retains the signal $\sigma aHf$, including its contribution from coordinate $i$. Varying the remaining scalar moves the vector along $v^{(i)}=\sigma\rho f_iH_{\cdot i}$, with $\norm{v^{(i)}}_\infty=\rho D^{-1/2}$.

Lemma~\ref{lem:taylor}, applied to $z\mapsto\Psi_f(\mu_0+zv^{(i)})$, gives constant, linear, and quadratic terms that cancel in conditional expectation because the moments match. The integral remainder is
\[
 R_i(z)=\frac{z^3}{2}\int_0^1(1-t)^2
      D_{v^{(i)}}^3\Psi_f(\mu_0+tzv^{(i)})\,dt.
\]
By \eqref{eq:magnitude-hyp} and the bound on $v^{(i)}$,
\[
 |R_i(z)|\le C B D^{-3/2}|z|^3
       (1+\norm{\mu_0}_\infty+\rho D^{-1/2}|z|)^k.
\]
\emph{Averaging the remainder.} We now average over uniform $f$. The following centering and subgaussian bounds are unconditional, not assertions for a fixed $f$. Write $\mu_0=\sigma\rho H A^{(i)}$, where
\[
 A_j^{(i)}=\begin{cases}
 f_j(m_0+\widetilde\xi_j),&j<i,\\
 m_0f_i,&j=i,\\
 f_jR_j,&j>i.
 \end{cases}
\]
These coordinates are independent, centered, and $1$-subgaussian. Indeed, $f_jR_j$ is standard normal for $j>i$, while for $j<i$,
\[
 \E e^{tf_j(m_0+\widetilde\xi_j)}
 =e^{v_0t^2/2}\cosh(m_0t)\le e^{t^2/2},
\]
since $\cosh u\le e^{u^2/2}$ and $m_0^2+v_0=1$. The same inequality handles $m_0f_i$.
Since the rows of $H$ have norm one, Lemma~\ref{lem:maxmoment} bounds the maximum moments of $\mu_0$.
Each $z\in\{\xi_i,\widetilde\xi_i\}$ is independent of $\mu_0$ and has bounded scalar moments, so
\[
 \begin{aligned}
 \E|R_i(z)|
 &\le C B D^{-3/2}\left[
   \E|z|^3\,\E(1+\norm{\mu_0}_\infty)^k
   +D^{-k/2}\E|z|^{k+3}\right]\\
 &\le C B D^{-3/2}(\log(2D))^{k/2}.
 \end{aligned}
\]
Summing the replacement errors gives \eqref{eq:magnitude}. Finally, $\sigma HD_fZ$ is standard normal conditional on $f$, with a distribution independent of $f$. This identifies the fully replaced vector with $\sigma aHf+\rho\sqrt{v_0}Z$, as stated.
\end{proof}

Sections~\ref{sec:ip} and~\ref{sec:ph} apply these tools with derivative bounds from verifier and circuit resources.

\section{The \texorpdfstring{$\IP$}{IP} bound}\label{sec:ip}

We first represent public-coin verifiers by average--maximum acceptance functions, fixing the input and accessible side-oracle bits so that only $(f,g)$ varies. We then prove family (i) of Theorem~\ref{thm:finite-common} for every function in that family. Smooth convex approximation gives derivative bounds for the comparisons along $\Ecal_{n,\sigma},\Pcal_{n,\sigma},\G_{n,\sigma},\Unif_n$. Comparing both signs with the uniform endpoint yields the desired gap, uniformly over side-oracle assignments.

\subsection{Verifiers as average--maximum circuits}\label{subsec:full-ip}
The following reduction lets us assume public coins.

\begin{theorem}[Goldwasser--Sipser, relativized]\label{ext:gs}
For every classical Boolean oracle $O$,
\[
 \IP^O=\AM[\poly]^O.
\]
Here $\AM[\poly]$ denotes the class of polynomial-round public-coin interactive proofs. In particular, every polynomial-round private-coin interactive proof has a polynomial-round public-coin simulation with polynomial resources and the standard constant completeness and soundness bounds.
\end{theorem}
The transformation~\cite[Section~4.2]{GS} uses random hashes and checks claimed transcripts and private tapes by simulating the old verifier. For each fixed classical oracle, these consistency checks take polynomial time with oracle access, so the hashing and counting argument relativizes (see also~\cite[Section~4.2]{BHNST}).

We put the public-coin verifier in a normal form in which all dependence on $(f,g)$ occurs in a final deterministic check. Each deterministic oracle-dependent verifier message moves into the next prover reply as a claim. After fresh coins $r_j$, the old exchange of $m_j$ and reply $a_j$ becomes a reply $(\widehat m_j,a_j)$, received before any later coins. The final check reconstructs the old verifier messages and requires $\widehat m_j=m_j$; a dummy reply carries a final claim if needed.

This preserves completeness. For soundness, an old-protocol prover follows the claimed transcript until the first inconsistency. Every accepting new transcript is consistent, so its acceptance probability is bounded by that of the simulated old strategy.

We enforce polynomial bounds on time, numbers of messages and queries, and all message, coin-string, and oracle-query lengths on every history, including malformed replies and arbitrary oracle inputs. Message lengths are prescribed in advance, with bounded replies encoded in fixed-width slots and malformed encodings rejected.

Dummy replies and fresh unused coins pad early termination and coin messages. Reconstructing the claimed messages takes polynomial time, so these changes preserve completeness and soundness with polynomial overhead. Here $b$ bounds prescribed message lengths after encoding and padding, as in~\cite[Lemma~4.2]{BHNST}.

\paragraph{From the final check to a convex function.}
We use the average--maximum framework of~\cite[Lemma~5.1]{AGH} and its adaptive-query formulation in~\cite[Lemma~4.2]{BHNST}, with the parameters of Definition~\ref{def:avgmax}. Since averages and maxima preserve convexity, we need only extend the final oracle check. For a fixed transcript and side oracle, this check is an adaptive decision tree on $(f,g)$. The next lemma constructs a convex extension from its accepting paths, together with a disjunction of path conditions that we will use for $\PH$ in Section~\ref{sec:ph}.

\begin{lemma}[Paths as clauses and affine penalties]\label{lem:paths}
Let $T:\{\pm1\}^m\to\{0,1\}$ have an adaptive decision tree of query depth at most $t$. For each accepting path $\gamma$, let $i_j$ be its query locations and $z_j$ the corresponding sign answers, and define
\[
 m_\gamma(x)=\sum_{j\in\gamma}\frac{1-z_jx_{i_j}}2.
\]
On Boolean inputs,
\begin{equation}\label{eq:paths}
 T(x)=\bigvee_{\gamma\ {\rm accepting}}
             \bigwedge_{j\in\gamma}[x_{i_j}=z_j]
     =\max\bigl\{0,\max_{\gamma\ {\rm accepting}}
                                   (1-m_\gamma(x))\bigr\}.
\end{equation}
The disjunctive normal form (DNF) has at most $2^t$ terms of width at most $t$. The maximum has at most $2^t+1$ affine leaves, each with gradient $\ell_1$ norm at most $t/2$, and defines a convex extension $\widetilde T$ satisfying
\[
 0\le\widetilde T(x)\le1\qquad(x\in[-1,1]^m).
\]
If there are no accepting paths, both expressions are zero.
\end{lemma}
\begin{proof}
On a Boolean input, the actual path is consistent and every other path has a mismatched answer. Thus each inconsistent path has a false clause and an affine score at most zero, while a consistent accepting path has score one. This proves \eqref{eq:paths}, including repeated or contradictory queries. The triangle inequality bounds each affine coefficient norm by $t/2$. On the cube, the mismatch penalties are nonnegative, so every score is at most one; the zero option gives the lower bound.
\end{proof}
Adaptive addresses are fixed once a path is specified. Side queries can be evaluated and their inconsistent branches discarded before applying the lemma. Thus its bounds are uniform over the fixed side bits.

\begin{lemma}[Public-coin verifiers as average--maximum circuits]\label{lem:am-compile}
Consider a verifier in the public-coin normal form above, with at most $r$ messages, each of a prescribed length at most $b$, and at most $q$ oracle queries in the final check. Fix the input and all oracle bits outside one target block. Its optimal acceptance probability, as a function of that block, has an average--maximum extension with
\[
 d\le r+1,\qquad B\le\max(b,q+1),\qquad \ell\le q/2.
\]
\end{lemma}
\begin{proof}
Unrolling the interaction gives a tree with an average over equally likely random strings at each verifier node and a maximum over possible replies at each prover node. A message slot of prescribed length $b_j\le b$ has $2^{b_j}\le2^b$ possible strings, so these gates have log fanin at most $b$. Invalid reply encodings lead to rejection. At each completed transcript, evaluating the fixed side oracle leaves an adaptive target-query tree of depth at most $q$. By Lemma~\ref{lem:paths}, its convex extension is a maximum of at most $2^q+1$ affine functions with coefficient norms at most $q/2$. This maximum adds one to the depth and has log fanin at most $q+1$. Averages and maxima preserve the range $[0,1]$ on the cube. The tree also preserves chronology, since a prover's maximum depends on past messages and every later fresh coin is averaged below that node. Backward induction therefore computes optimal acceptance over all prover strategies.
\end{proof}
For example, a protocol with two rounds, each consisting of a coin message and a reply, has value
\[
 \E_{r_1}\max_{a_1}\E_{r_2}\max_{a_2}
 T(f,g;r_1,a_1,r_2,a_2).
\]
The first reply can depend on $r_1$, and the second can also depend on $r_2$. Interaction controls depth $d$; message lengths and the number of terminal queries control log fanin $B$; and the terminal query count controls leaf sensitivity $\ell$. Our estimates use only these three parameters, even when the tree is exponentially large.

\subsection{Softmax smoothing and derivative bounds}
The count-repair comparison of Section~\ref{sec:coupling} uses convexity; the Gaussian comparisons also require derivatives. Replacing maximum gates by the smooth log-sum-exponential preserves convexity with controlled error. We use the smoothed functions to analyze expectations under the original distributions.

The derivative bounds $L_1,L_2,L_3$ from \eqref{eq:derivative-budgets} control Lipschitz errors, covariance comparison, and magnitude replacement, respectively. Larger $\lambda$ improves the approximation but increases higher derivatives. We use the softmax induction of \cite[Lemma~6.1]{BHNST}, separating log fanin $B$ from leaf sensitivity $\ell$, and extend its estimate to order three.

\begin{lemma}[Smoothing average--maximum circuits]\label{lem:softmax}
For a $(d,B,\ell)$ average--maximum circuit $C$ and $\lambda>0$, replacing every maximum by
\[
 S_\lambda(z_1,\ldots,z_k)=\lambda^{-1}\log\sum_i e^{\lambda z_i}
\]
gives a smooth convex $F$ with approximation error
\begin{equation}\label{eq:smooth-approx}
 0\le F-C\le\errsm:=dB\log2/\lambda.
\end{equation}
Its derivative bounds are
\begin{align}
 L_1(F)&\le\ell,\label{eq:smooth-one}\\
 L_2(F)&\le2\lambda d\ell^2,\label{eq:smooth-two}\\
 L_3(F)&\le(6d^2+2d)\lambda^2\ell^3.\label{eq:smooth-three}
\end{align}
All estimates hold on $\R^m$.
\end{lemma}
\begin{proof}
At a gate with $k\le2^B$ children,
\[
 0\le S_\lambda(z)-\max_i z_i\le\frac{\log k}{\lambda}
 \le\frac{B\log2}{\lambda}.
\]
Adding the same constant to all child values adds that constant to $S_\lambda$. Thus errors accumulate along the depth. If each smoothed child value $\widetilde u_i$ differs from its original value $u_i$ by a number in $[0,e]$, this property and monotonicity give
\[
 \max_i u_i\le S_\lambda(\widetilde u)
       \le\max_i u_i+e+\frac{\log k}{\lambda}.
\]
Averages also propagate at most the largest child error, proving \eqref{eq:smooth-approx}. For the derivative bounds, consider a softmax node with smooth child functions $u_i$ and write
\[
 p_i=\frac{e^{\lambda u_i}}{\sum_j e^{\lambda u_j}},\quad
 g_i=\nabla u_i,\quad B_i=\nabla^2u_i,\quad \bar g=\sum_i p_i g_i.
\]
All quantities are evaluated at the same input, and $\E_p$ denotes averaging over the child index with weights $p_i$. Differentiation gives
\[
 \nabla S_\lambda=\bar g,\qquad
 \nabla^2 S_\lambda=\sum_i p_iB_i+
                  \lambda\sum_i p_i(g_i-\bar g)(g_i-\bar g)^T.
\]
The child Hessians are positive semidefinite by induction, as is the covariance term, so the smoothed function is convex. The gradient bound follows from $\|\bar g\|_1\le\E_p\|g_i\|_1\le\ell$. To bound the Hessian, we expand the covariance before taking its entrywise norm and obtain
\[
 \begin{aligned}
 \operatorname{Cov}_p(g_i)&=\E_p[g_i g_i^T]-\bar g\bar g^T,\\
 \|\operatorname{Cov}_p(g_i)\|_1
 &\le\E_p\|g_i\|_1^2+\|\bar g\|_1^2\le2\ell^2.
 \end{aligned}
\]
Thus, if $L_{2,r}$ bounds the Hessian norm through height $r$, then $L_{2,r}\le L_{2,r-1}+2\lambda\ell^2$. The affine leaves have zero Hessian, giving $L_{2,r}\le2\lambda r\ell^2$.

For the third derivative, put $h_i=g_i-\bar g$, $\bar B=\E_p B_i$, and $T_i=\nabla^3u_i$. The identity $\partial_c p_i=\lambda p_i h_{i,c}$ gives the differentiation rule
\[
 \partial_c\E_p A_i
 =\E_p[\partial_c A_i]+\lambda\E_p[A_i h_{i,c}]
\]
for a scalar or tensor entry $A_i$. Applying this rule to the gradient and Hessian gives
\[
 \begin{aligned}
 \partial_{abc}S_\lambda
 ={}&\E_p T_{i,abc}
     +\lambda^2\E_p[h_{i,a}h_{i,b}h_{i,c}]\\
 &+\lambda\E_p\bigl[
       (B_i-\bar B)_{ab}h_{i,c}
       +(B_i-\bar B)_{ac}h_{i,b}
       +(B_i-\bar B)_{bc}h_{i,a}\bigr].
 \end{aligned}
\]
In each of the three mixed terms, $\E_p h_i=0$ cancels the contribution of $\bar B$. The entrywise identity $\|B\otimes h\|_1=\|B\|_1\|h\|_1$ and the bound $\|h_i\|_1\le2\ell$ then give a norm bound of $2\ell L_{2,r-1}$ for each term. The centered cubic term has norm at most $8\ell^3$. Thus, writing $L_{3,r}$ for the third-derivative bound through height $r$, we obtain
\[
 L_{3,r}\le L_{3,r-1}+6\lambda\ell L_{2,r-1}+8\lambda^2\ell^3
 \le L_{3,r-1}+[12(r-1)+8]\lambda^2\ell^3.
\]
At each node the softmax weights average over children, so the recurrence accumulates over tree height. The leaves have no second or third derivatives, and summing from height one to $d$ gives $6d(d-1)+8d=6d^2+2d$, as required. At average gates, the triangle inequality preserves all three derivative bounds.
\end{proof}

\subsection{\texorpdfstring{Comparing $\G_{n,\sigma}$ and $\Unif_n$ for convex tests}{Comparing G(n,sigma) and U(n) for convex tests}}
The distributions $\G_{n,\sigma}$ and $\Unif_n$ are signs of Gaussian vectors with covariances $\Sigma_{n,\sigma}$ and $I$. The following imported theorem handles the discontinuous sign map.

\begin{external}[Shifted Gaussian-sign comparison~{\cite[Theorem~6.4]{BHNST}}]\label{ext:shifted-convex}
Let $Q:\R^m\to\R$ be smooth and convex. For the cube derivative norms in \eqref{eq:derivative-budgets}, assume
\[
 L_1(Q)\le L,\qquad L_2(Q)\le K,\qquad L\le m.
\]
Let $V\sim\Normal(0,\Sigma)$ and $Z\sim\Normal(0,I_m)$ be independent. Assume that $\Sigma$ has unit diagonal, and write $\beta=\max_{i\ne j}|\Sigma_{ij}|$. For every fixed shift $u\in\R^m$ and $0<\delta\le1/2$,
\begin{equation}\label{eq:shifted-convex-input}
 \begin{aligned}
  \left|\E Q\!\left(\sgn\!\left(u+\frac{Z+V}{\sqrt2}\right)\right)
          -\E Q(\sgn(u+Z))\right|\le400L\delta\sqrt{\log(2m)}
                   +\frac{\beta K}{4\delta^2}+\frac4{m^{48}}.
 \end{aligned}
\end{equation}
\end{external}
The bound is uniform in $u$, so we may condition on an independent random shift and then average. Following~\cite[proof of Corollary~6.6]{BHNST}, we apply it in three steps using the endpoints' fixed independent-noise component.

\begin{proposition}[Gaussian-sign comparison for convex tests]
\label{prop:convex-gaussian}
Let $F:\R^{2D}\to\R$ be smooth and convex, with $L_1(F)\le L_1$ and $L_2(F)\le L_2$ on $[-1,1]^{2D}$. For either $\sigma\in\{+1,-1\}$ and every $0<\delta\le1/2$,
\begin{equation}\label{eq:convex-gaussian}
 \left|\E_{\G_{n,\sigma}}F-\E_{\Unif_n}F\right|
 \le C\left[
 L_1\delta\sqrt{\log(2D)}+D^{-1/2}L_2\delta^{-2}+L_1D^{-10}\right].
\end{equation}
\end{proposition}
\begin{proof}
Write $\Sigma=\Sigma_{n,\sigma}$. To separate off independent noise with covariance $I/8$, note that the smallest eigenvalue is $1-\rho=1-\pi/4>1/8$. Hence
\[
 \widetilde\Sigma=\frac87\left(\Sigma-\frac18 I\right)
\]
is a positive definite covariance matrix with unit diagonal and off-diagonal entries of magnitude at most $(8\rho/7)D^{-1/2}$. Take independent $V\sim\Normal(0,\widetilde\Sigma)$ and $Z\sim\Normal(0,I)$, and set
\[
 V_j=2^{-j/2}Z+\sqrt{1-2^{-j}}\,V,\qquad j=0,1,2,3.
\]
Each step halves the independent covariance. Thus $V_0$ has covariance $I$, while $V_3$ has covariance $I/8+(7/8)\widetilde\Sigma=\Sigma$. Their sign vectors have distributions $\Unif_n$ and $\G_{n,\sigma}$.

Let $V'$ be an independent copy of $V$, independent of $(V,Z)$. For $j=0,1,2$, put $u_j=\sqrt{2^j-1}\,V$. The common-shift representation is
\[
 \sgn V_j\stackrel d=\sgn(u_j+Z),\qquad
 \sgn V_{j+1}\stackrel d=
 \sgn\!\left(u_j+\frac{Z+V'}{\sqrt2}\right).
\]
These are identities of marginal laws. Positive rescaling gives the first, and the second follows from
\[
 \begin{gathered}
 \operatorname{Cov}\!\left(
  2^{-(j+1)/2}(Z+V')+\sqrt{1-2^{-j}}\,V\right)\\
 =2^{-(j+1)}I+(1-2^{-(j+1)})\widetilde\Sigma
 =\operatorname{Cov}(V_{j+1}).
 \end{gathered}
\]
Conditional on $V$, both noises retain their laws and are independent of the common shift. We can therefore apply \eqref{eq:shifted-convex-input} and average over $V$.

If $L_1=0$, $F$ is constant on the cube and the claim is immediate. Otherwise, apply \eqref{eq:shifted-convex-input} with $Q=F/L_1$, $m=2D$, $L=1$, $K=L_2/L_1$, and $\beta\le(8\rho/7)D^{-1/2}$, so that $L\le m$. Multiplication by $L_1$ bounds the cost of each step by
\[
 C\left[L_1\delta\sqrt{\log(2D)}
        +D^{-1/2}L_2\delta^{-2}+L_1D^{-48}\right].
\]
Summing the three steps and using $D^{-48}\le D^{-10}$ proves the bound.
\end{proof}

\subsection{\texorpdfstring{Comparing $\Pcal_{n,\sigma}$ and $\G_{n,\sigma}$ for convex tests}{Comparing P(n,sigma) and G(n,sigma) for convex tests}}
Lemma~\ref{lem:magnitude} requires smooth functions of the vector before signs are taken. Softmax alone does not make the composition with sign continuous. We therefore replace sign by a smooth function agreeing with it outside $[-\zeta,\zeta]$, then bound the resulting error at both endpoints.

A union bound on coordinates entering that interval would lose a factor proportional to $D$. The conditional-mean comparison from Section~\ref{sec:coupling} avoids this loss by bounding changes in plus probabilities relative to the less likely sign. The relative bound accommodates arbitrarily biased signs after the Gaussian means are fixed.

\begin{lemma}[Nearby product sign probabilities]\label{lem:product-bias}
Let $S,S'\in\{\pm1\}^m$ have product distributions with plus probabilities $p_i,p_i'$, where $0<p_i<1$. For every convex $L_1$-Lipschitz function $F$ on $[-1,1]^m$,
\[
 |\E F(S)-\E F(S')|
 \le 2L_1\max_i\frac{|p_i-p_i'|}{\min(p_i,1-p_i)}.
\]
\end{lemma}
\begin{proof}
Couple the coordinates with independent uniforms, using thresholds $p_i$ and $p_i'$. To apply Lemma~\ref{lem:barycenter}, we bound the changes in conditional means. If $p_i'\ge p_i$, only a minus sign of $S_i$ can become plus, while in the reverse direction only a plus sign of $S_i'$ can become minus. The corresponding changes have magnitudes
\[
 \frac{2(p_i'-p_i)}{1-p_i},\qquad
 \frac{2(p_i'-p_i)}{p_i'}.
\]
Both denominators are at least $\min(p_i,1-p_i)$. If $p_i'<p_i$, the denominators are $p_i$ and $1-p_i'$, giving the same bound. By independence, these bounds still hold when conditioning on the whole corresponding sign vector, so Lemma~\ref{lem:barycenter} proves the claim.
\end{proof}

\paragraph{A smooth sign cutoff.}
Choose a smooth odd nondecreasing cutoff $\chi_\zeta$, valued in $[-1,1]$, equal to sign off $[-\zeta,\zeta]$, and satisfying
\[
 \|\chi_\zeta^{(j)}\|_\infty\le C_j\zeta^{-j},\qquad j=1,2,3.
\]
Convexity gives both bounds needed to remove the cutoff, a lower bound by conditioning on the original signs and an upper bound by rounding the smooth values to signs.

\begin{lemma}[Removing the sign cutoff]\label{lem:rounding}
Let $V=(V_j)_{j=1}^m$ have independent coordinates $V_j\sim\Normal(\mu_j,s^2)$, where $s>0$ is fixed and $\max_j|\mu_j|\le B_\mu$. If $\zeta(1+B_\mu)$ is sufficiently small, then every convex $L_1$-Lipschitz function $F$ on $[-1,1]^m$ satisfies
\[
 |\E F(\chi_\zeta(V))-\E F(\sgn V)|
 \le C_s L_1\zeta(1+B_\mu).
\]
\end{lemma}
\begin{proof}
Put $p_j=\Pp(V_j>0)$. By the normal density and \eqref{eq:mills},
\[
 \begin{aligned}
 \Pp(|V_j|\le\zeta)
 &\le(2\zeta/s)\varphi(\mu_j/s)e^{|\mu_j|\zeta/s^2},\\
 \min(p_j,1-p_j)&\ge
 \frac{c\,\varphi(\mu_j/s)}{1+|\mu_j|/s}.
 \end{aligned}
\]
The exponential is bounded under the stated smallness condition, so
\begin{equation}\label{eq:relative-small}
 \epsilon:=\max_j
 \frac{\Pp(|V_j|\le\zeta)}{\min(p_j,1-p_j)}
 \le C_s\zeta(1+B_\mu).
\end{equation}

For the lower bound, let $S=\sgn V$. Coordinate independence gives
$\E[\chi_\zeta(V_j)\mid S]=\E[\chi_\zeta(V_j)\mid S_j]$.
Since the cutoff differs from sign only in $[-\zeta,\zeta]$ and by at most two,
\[
 \begin{aligned}
 \norm{\E[\chi_\zeta(V)\mid S]-S}_\infty&\le2\epsilon,\\
 \E F(\chi_\zeta(V))&\ge\E F(S)-2L_1\epsilon,
 \end{aligned}
\]
where the second inequality uses conditional Jensen and the Lipschitz bound.

For the upper bound, choose the signs $S_j'$ independently conditional on $V$, with
\[
 \Pp(S_j'=+1\mid V)=\frac{1+\chi_\zeta(V_j)}2,
 \qquad \E[S'\mid V]=\chi_\zeta(V).
\]
Each $S_j'$ is generated from $V_j$ and its own independent uniform variable, so $S'$ has a product law. Its plus probabilities satisfy
\[
 |p_j'-p_j|
 =\tfrac12|\E\chi_\zeta(V_j)-\E\sgn(V_j)|
 \le\Pp(|V_j|\le\zeta).
\]
Jensen and Lemma~\ref{lem:product-bias} therefore give
\[
 \E F(\chi_\zeta(V))\le\E F(S')
 \le\E F(S)+2L_1\epsilon.
\]
Together with \eqref{eq:relative-small}, this proves the claim.
\end{proof}

\paragraph{Applying magnitude replacement.}
Magnitude replacement uses third derivatives; cutoff removal uses convexity and the Lipschitz bound. Only $F$ needs to be convex, not its composition with the cutoff.

\begin{proposition}[Convex magnitude replacement]\label{prop:ip-replace}
Let $F$ be smooth and convex on a neighborhood of $[-1,1]^{2D}$, with the derivative bounds $L_1,L_2,L_3$ of \eqref{eq:derivative-budgets} holding there. For $0<\zeta\le1$ and $\zeta\sqrt{\log(2D)}$ sufficiently small,
\begin{align}
 |\E_{\G_{n,\sigma}}F-\E_{\Pcal_{n,\sigma}}F|
 \le C\bigl[&
 L_1\zeta\sqrt{\log(2D)}
 +(L_1+L_2+L_3)\zeta^{-3}D^{-1/2}
 +L_1D^{-10}\bigr].\label{eq:ip-replace}
\end{align}
\end{proposition}
\begin{proof}
\emph{Replacement cost.} For independent standard normal $Y$, define
\[
 \Psi_f(\mu)=\E_Y F(f,\chi_\zeta(\mu+sY)).
\]
Fix $f$, with noise scale $s$ as in \eqref{eq:constants}, and write $\chi_j^{(r)}=\chi_\zeta^{(r)}(\mu_j+sY_j)$. Evaluate the partial derivatives of $F$ at $(f,\chi_\zeta(\mu+sY))$. The chain rule gives
\[
 \begin{aligned}
 D_v^3\Psi_f(\mu)=\E_Y\bigg[
 \sum_{j,k,l}F_{jkl}\chi_j'\chi_k'\chi_l'v_jv_kv_l
 +3\sum_{j,k}F_{jk}\chi_j''\chi_k'v_j^2v_k
 +\sum_j F_j\chi_j'''v_j^3\bigg].
 \end{aligned}
\]
All indices refer to the $g$ coordinates, including repetitions. Since the cutoff keeps the argument of $F$ inside the cube, the bounded derivatives of $F$ and of the cutoff justify differentiation under $\E_Y$ and give
\[
 |D_v^3\Psi_f(\mu)|
 \le C(L_1+L_2+L_3)\zeta^{-3}\|v\|_\infty^3.
\]
This is \eqref{eq:magnitude-hyp} with $k=0$ and $B=C(L_1+L_2+L_3)\zeta^{-3}$. Lemma~\ref{lem:magnitude} gives the middle term of \eqref{eq:ip-replace}, with expectations averaged over uniform $f$. At the fully replaced endpoint, adding $sY$ combines the independent noise variances to $\rho^2v_0+s^2=b^2$, identifying $\Pcal_{n,\sigma}$ once the cutoff is removed. The original endpoint similarly gives $\G_{n,\sigma}$.

\emph{Endpoint errors.} Conditional on $f$ and the latent mean $\mu$, the coordinates of the noise $sY$ remain independent, as required by Lemma~\ref{lem:rounding}; the coordinates of $\mu$ itself need not be independent. At the two endpoints, $\mu=\sigma\rho HX$ or $\mu=\sigma aHf+\rho\sqrt{v_0}Z$. The first has Gaussian marginals of variance $\rho^2$. For the second, averaging over uniform $f$ and independent $Z$ gives
\[
 \E e^{t\mu_j}\le
 \exp\!\left(\tfrac12(a^2+\rho^2v_0)t^2\right)
 =\exp(\rho^2t^2/2),
\]
using the subgaussian row bound for $Hf$ and $a^2+\rho^2v_0=\rho^2$. Lemma~\ref{lem:maxmoment} therefore gives $\norm\mu_\infty\le C\sqrt{\log(2D)}$ except on an event of probability at most $D^{-10}$. Under this norm bound, Lemma~\ref{lem:rounding} bounds the cutoff error for $F(f,\cdot)$ by $CL_1\zeta\sqrt{\log(2D)}$. On the exceptional event both arguments lie in the same cube, so their $F$-values differ by at most $2L_1$. Applying these estimates at both endpoints and adding the replacement cost proves the proposition.
\end{proof}

\subsection{Combining the comparisons}
Table~\ref{tab:error-budget} collects the costs, using $\log(2D)=\Theta(n)$. The parameters are softmax scale $\lambda$, Gaussian sign-smoothing width $\delta$, magnitude-replacement cutoff width $\zeta$, and criticality $\Lambda$ from Definition~\ref{def:criticality}. We choose $\delta=\zeta$ after balancing their distinct costs.

\begin{table}[!htb]
\centering
\normalsize
\begingroup
\renewcommand{\arraystretch}{1.0}
\setlength{\jot}{1.5pt}
\begin{tabular}{@{}
  >{\raggedright\arraybackslash}m{0.22\textwidth}
  @{\hspace{0.03\textwidth}}
  >{\raggedright\arraybackslash}m{0.49\textwidth}
  @{\hspace{0.03\textwidth}}
  >{\raggedright\arraybackslash}m{0.23\textwidth}@{}}
\toprule
\textbf{Step} & \textbf{Convex/$\IP$ cost} & \textbf{Boolean/$\PH$ cost}\\
\specialrule{\lightrulewidth}{0.25em}{0.4em}
\textbf{Count repair}\newline
{\normalsize (Prop.~\ref{prop:repair})}
& $\displaystyle \frac{L_1}{M}+\frac{\osc(F)}{M^2}$
& $\displaystyle \frac\Lambda M+\frac1{M^2}$\\
\specialrule{0.3pt}{0.4em}{0.4em}
\textbf{Magnitude}\newline \textbf{replacement}\newline
{\normalsize (Props.~\ref{prop:ip-replace}, \ref{prop:ph-replace})}
& $\begin{aligned}
 &L_1\zeta\sqrt n\\
 &\quad{}+D^{-1/2}(L_1+L_2+L_3)\zeta^{-3}\\
 &\quad{}+L_1D^{-10}
 \end{aligned}$
& $D^{-1/2}\Lambda^3n^{3/2}$\\
\specialrule{0.3pt}{0.4em}{0.4em}
\textbf{Gaussian}\newline \textbf{comparison}\newline
{\normalsize (Props.~\ref{prop:convex-gaussian}, \ref{prop:ph-gaussian})}
& $\begin{aligned}
 &L_1\delta\sqrt n\\
 &\quad{}+D^{-1/2}L_2\delta^{-2}\\
 &\quad{}+L_1D^{-10}
 \end{aligned}$
& $D^{-1/2}\Lambda^2n$\\
\specialrule{0.3pt}{0.4em}{0.4em}
\textbf{Acceptance}\newline \textbf{smoothing}\newline
{\normalsize (Lemma~\ref{lem:softmax})}
& $dB/\lambda$
& $0$\newline {\normalsize (exact Gaussian averaging)}\\
\addlinespace[0.3em]
\bottomrule
\end{tabular}
\endgroup
\caption{Error terms for Theorem~\ref{thm:finite-common}, all in $O(\cdot)$ with constants depending only on fixed construction parameters. The first three rows compare $\Ecal_{n,\sigma}$ to uniform; the last removes the softmax approximation. The $\PH$ column previews Section~\ref{sec:ph}.}
\label{tab:error-budget}
\end{table}

\paragraph{The master bound.}
Let $C$ be an average--maximum acceptance function and $F$ its softmax approximation. Since $0\le C\le1$ on Boolean inputs, $\osc_{\{\pm1\}^{2D}}F\le1+\errsm$. Applying the three comparisons to $F$ and accounting for its approximation error gives
\[
 \begin{aligned}
 |\E_{\Ecal_{n,\sigma}}C-\E_{\Unif_n}C|
 \le{}&2\errsm
       +|\E_{\Ecal_{n,\sigma}}F-\E_{\Pcal_{n,\sigma}}F|\\
      &+|\E_{\Pcal_{n,\sigma}}F-\E_{\G_{n,\sigma}}F|
       +|\E_{\G_{n,\sigma}}F-\E_{\Unif_n}F|.
 \end{aligned}
\]
Substituting Propositions~\ref{prop:repair}, \ref{prop:ip-replace}, and \ref{prop:convex-gaussian} gives, whenever their smoothing conditions hold,
\begin{equation}\label{eq:ip-master-bound}
 \begin{aligned}
 \left|\E_{\Ecal_{n,\sigma}}C-\E_{\Unif_n}C\right|
 \le C_0\bigg[&
 \frac{dB}{\lambda}+\frac{\ell}{M}
 +\frac{1+dB/\lambda}{M^2}+\ell(\delta+\zeta)\sqrt n\\
 &+D^{-1/2}\left(L_2\delta^{-2}
       +(L_1+L_2+L_3)\zeta^{-3}\right)+\ell D^{-10}\bigg].
 \end{aligned}
\end{equation}
\paragraph{Parameter choice.}
A larger $\lambda$ reduces the softmax error and increases derivative bounds; smaller cutoff widths reduce endpoint errors and increase replacement costs. Fix a polynomial $q=q(n)\ge n+1$ bounding $d,B,\ell$ as in Theorem~\ref{thm:finite-common}. Balancing the leading approximation, cutoff-removal, and magnitude-replacement costs for $\lambda\ge1$ gives
\[
 \frac{q^2}{\lambda}\asymp q\zeta\sqrt n
 \asymp D^{-1/2}q^5\lambda^2\zeta^{-3}
 \quad\Longrightarrow\quad
 e\asymp q^2n^{1/4}D^{-1/12},
\]
where $e$ is their common scale. We take $\delta=\zeta$. The covariance cost is then smaller, as checked below.

\begin{proof}[Proof of Theorem~\ref{thm:finite-common}, average--maximum case]
Write $\nu=n^{1/4}$ and choose $\lambda=M/\nu$, $\delta=\zeta=q/(M\nu)$. For all sufficiently large lengths $\lambda\ge1$, so Lemma~\ref{lem:softmax} gives
\begin{equation}\label{eq:common-smoothing}
 L_1\le q,\qquad L_2=O(q^3\lambda),\qquad
 L_1+L_2+L_3=O(q^5\lambda^2).
\end{equation}
The softmax test is smooth and convex on all of $\R^{2D}$. Since $q(n)$ is a fixed polynomial, eventually $\delta\le1/2$ and $\zeta\sqrt{\log(2D)}=O(q\nu/M)\to0$, as required. Using $D^{-1/2}=O(M^{-6})$, the derivative costs satisfy
\[
 \begin{aligned}
 D^{-1/2}(L_1+L_2+L_3)\zeta^{-3}
 &\le C M^{-6}\frac{q^5M^2}{\nu^2}
                  \frac{M^3\nu^3}{q^3}=O(q^2\nu/M),\\
 D^{-1/2}L_2\delta^{-2}
 &\le C M^{-6}\frac{q^3M}{\nu}
                  \frac{M^2\nu^2}{q^2}=O(q\nu/M^3).
 \end{aligned}
\]
The repair, softmax, cutoff, and tail costs are also $O(q^2\nu/M)$, so \eqref{eq:ip-master-bound} has this bound for each sign. Comparing both signs to the common uniform endpoint and using $M=\Theta(2^{n/12})$ proves family (i), uniformly over side-oracle assignments.
\end{proof}

\section{The \texorpdfstring{$\PH$}{PH} bound}\label{sec:ph}

We prove the Boolean-test case of Theorem~\ref{thm:finite-common} and apply it to $\PH$ predicates. Fixing the accessible side-oracle bits gives a constant-depth circuit of size $\exp(\poly(n))$. Its restriction-stable criticality yields weighted derivative bounds; Gaussian averaging then compares the product and Gaussian-sign laws with uniform input. The count-repair estimate completes the argument.

\subsection{\texorpdfstring{$\PH$}{PH} circuits and restriction-stable criticality}\label{subsec:ph-criticality}
\begin{lemma}[Circuit representation of $\PH$ predicates]
\label{lem:ph-compile}
Fix a polynomial-clocked oracle predicate with $k$ quantified blocks, where $k$ is constant, on input $1^n$. Fix every accessible oracle bit outside the target block. Its value as a function of $(f,g)$ is computed by an AND/OR circuit of depth at most $k+2$ and size $\exp(\poly(n))$. These bounds are independent of the values of the fixed bits.
\end{lemma}
\begin{proof}
Write the predicate as
\[
 Q_1 y_1\in\{0,1\}^{b_1(n)}\ \cdots\
 Q_k y_k\in\{0,1\}^{b_k(n)}\
 R^O(1^n,y_1,\ldots,y_k),
\]
where $R^O$ is a deterministic polynomial-time oracle computation. All block lengths and the global terminal clock are bounded by a fixed polynomial. To put a fixed-alternation machine in this form, pad each block's sequence of binary choices and let the terminal simulation consume them in order, including its intervening deterministic oracle computations.

For each quantified tuple, unrolling the terminal computation with the side-oracle answers fixed gives a target-query decision tree of depth at most $q(n)=\poly(n)$. Lemma~\ref{lem:paths} gives its DNF, allowing adaptive addresses and repeated queries.

This DNF has at most $2^{q(n)}$ terms, each with at most $q(n)$ literals, and depth two. Each existential quantifier contributes an OR and each universal quantifier an AND over its strings. There are at most $2^{\sum_i b_i(n)}$ tuple copies, so the total size, including the $2D$ target inputs, is bounded by
\[
 O\!\left(2D+
  2^{\sum_i b_i(n)}(q(n)+1)2^{q(n)}\right)
 =\exp(\poly(n)).
\]
There are at most $k+2$ gate layers. Fixing side answers does not increase either bound.
\end{proof}

We next verify the criticality condition from Definition~\ref{def:criticality}. The bound must hold after arbitrary restrictions, since repair fixes the outer data and magnitude replacement fixes $f$. We use Rossman's criticality theorem for regular formulas and its bounded-depth circuit consequence~\cite[Theorem~3 and Section~1.2]{Rossman19}.

\begin{lemma}[Criticality of bounded-depth circuits]
\label{ext:circuit-criticality}\label{lem:critical-envelope}
Every AND/OR circuit of depth at most a fixed $d_{\rm circ}\ge2$ and total size $S_{\rm tot}$, including inputs, with negations only at the inputs, satisfies a restriction-stable criticality bound through order three with
\begin{equation}\label{eq:lambda-envelope}
 \Lambda\le c_{d_{\rm circ}}(1+\log S_{\rm tot})^{d_{\rm circ}-1}.
\end{equation}
\end{lemma}
\begin{proof}
Fix a deterministic restriction $\tau$ and put $h=d_{\rm circ}-1$. The circuit $C|_\tau$ still has depth at most $h+1$ and size at most $S_{\rm tot}$. The conversion in~\cite[Section~1.2]{Rossman19}, with fanin-one padding, gives an equivalent regular formula of depth $h+1$. Its number of depth-one subformulas satisfies $1\le s_{\rm reg}\le(S_{\rm tot}+1)^{O(d_{\rm circ})}$, hence $\ln s_{\rm reg}=O_{d_{\rm circ}}(1+\log S_{\rm tot})$. Lemma~\ref{lem:criticality-correction} in Appendix~\ref{app:criticality} therefore gives criticality at most $\Lambda=c_{d_{\rm circ}}(1+\log S_{\rm tot})^h$, with $c_{d_{\rm circ}}\ge1$ independent of $\tau$. Thus $\Pp_{\eta\sim\Res_t}[\DT(C|_{\tau,\eta})\ge r]\le(t\Lambda)^r$ for $t\in[0,1]$ and $r=1,2,3$, uniformly in $\tau$.
\end{proof}
Together with Lemma~\ref{lem:ph-compile}, this gives $\Lambda=\poly(n)$ for each fixed $\PH$ resource bound, uniformly over the fixed side-oracle bits and all further restrictions of the target inputs. The rest of this section uses only the restriction-stable bound from Definition~\ref{def:criticality}.

\subsection{From criticality to weighted derivatives}
We bound derivative sums by viewing each mixed derivative as a coefficient of a randomly restricted function. First, extend the Boolean test by averaging over independent signs with prescribed means.

\paragraph{Multilinear extension.}
For $C:\{\pm1\}^m\to\{0,1\}$, define
\begin{equation}\label{eq:multilinear}
 F(x)=\sum_{z\in\{\pm1\}^m} C(z)\prod_{i=1}^m\frac{1+x_i z_i}{2}.
\end{equation}
For $x\in[-1,1]^m$, the product in each summand is the probability of $z$ under independent signs of means $x_i$, so $F(x)=\E_x C(X)$. In particular, $F$ agrees with $C$ on Boolean inputs.

Here $F$ denotes the multilinear extension, which need not be convex; it replaces the softmax extension of Section~\ref{sec:ip}.

Subscripts $x$ or $\theta$ on $\E,\Pp$ denote this product sign measure. A sign with plus probability $p_i$ has mean $\theta_i=2p_i-1$. Here $m$ may be $2D$, or $D$ after $f$ is fixed. For $A\subseteq[m]$, write $\partial_A=\prod_{i\in A}\partial_i$. For a single coordinate,
\[
 \partial_iF(x)=\tfrac12\bigl(F(x|_{x_i=1})-F(x|_{x_i=-1})\bigr).
\]
Iterating this identity explains the discrete differences below. Multilinearity makes $\partial_A F$ independent of the coordinates in $A$.

Fixing the complementary coordinates turns $\partial_A F$ into the coefficient of the product of all $|A|$ remaining variables. If nonzero, it forces the restricted function to have degree $|A|$, and hence decision-tree depth at least $|A|$. We will sample the free set so that the restriction bound controls the entire derivative sum, without paying separately for each subset.

The bias weights $w_i=1-|\theta_i|$, twice the smaller sign probability, measure how much mass can be left free while preserving the bias upon fair completion. They also occur in derivatives of Gaussian sign means. They are distinct from the integer bin weights in Section~\ref{sec:laws}.
\begin{lemma}[Weighted derivatives]\label{lem:weighted}
Let $C:\{\pm1\}^m\to\{0,1\}$ satisfy Definition~\ref{def:criticality}, and let $F$ be its multilinear extension. For every $\theta\in[-1,1]^m$, put $w_i=1-|\theta_i|$. Then, for $r=1,2,3$,
\begin{align}
 \sum_{|A|=r}\left(\prod_{i\in A}w_i\right)
      \E_\theta|\partial_A F(X)|&\le\Lambda^r/2,\label{eq:weighted-mean}\\
 \sum_{|A|=r}\left(\prod_{i\in A}w_i\right)
      |\partial_A F(\theta)|&\le\Lambda^r/2.\label{eq:weighted-point}
\end{align}
Here $X$ has independent sign coordinates of means $\theta_i$. Both bounds also hold for the multilinear extension of any deterministic restriction of $C$, with the same $\Lambda$.
\end{lemma}
\begin{proof}
Fix $r\in\{1,2,3\}$ and $\theta$. For $0<t<1$, leave coordinate $i$ free with probability $tw_i$, subtracting half this mass from each of its original sign probabilities. For the resulting product restriction $\eta_t$, the probabilities $s_i,a_i,b_i$ of being free, fixed to $+1$, and fixed to $-1$, respectively, are
\[
 s_i=tw_i,\qquad a_i=\frac{1+\theta_i-tw_i}{2},\qquad
 b_i=\frac{1-\theta_i-tw_i}{2}.
\]
They are nonnegative because $tw_i\le1-|\theta_i|$. Filling each free coordinate with a fair sign restores plus probability $a_i+s_i/2=(1+\theta_i)/2$, so it restores the original product measure. In Lemma~\ref{lem:biased-transfer}, $a_i-b_i=\theta_i$ and
\[
 \frac{s_i}{1-|a_i-b_i|}=\frac{tw_i}{w_i}=t
 \qquad\text{when }w_i>0.
\]
When $w_i=0$, the coordinate is deterministic and the ratio is zero by convention. Hence $t_*\le t$, and the lemma gives $\Pp[\DT(C|_{\eta_t})\ge r]\le(t\Lambda)^r$.

We count a derivative only when its indices are exactly the free set. Let $A=A(\eta_t)$ and define
\[
 T_r(\eta_t)=\one_{\{|A|=r\}}
                 |\partial_A F((\eta_t)_{A^c})|.
\]
For $|A|=r$ and fixed complementary signs $z$, the coefficient is
\[
 \partial_A F(z)=2^{-r}\sum_{\varepsilon\in\{\pm1\}^A}
       C(\varepsilon,z)\prod_{i\in A}\varepsilon_i.
\]
The coefficient has magnitude at most $1/2$ and, when nonzero, gives degree $r$ on the free variables. Since polynomial degree is at most decision-tree depth,
\[
 T_r(\eta_t)\le\tfrac12\one_{\{\DT(C|_{\eta_t})\ge r\}},
 \qquad
 \E T_r(\eta_t)\le\tfrac12(t\Lambda)^r.
\]
For a fixed $r$-set $A$, the probability of leaving exactly $A$ free is $t^r\prod_{i\in A}w_i\prod_{j\notin A}(1-tw_j)$. Conditional on this event, the other signs are independent with means
\[
 \theta_j^{(t)}=\frac{a_j-b_j}{1-s_j}
              =\frac{\theta_j}{1-tw_j},\qquad j\notin A.
\]
The events specifying the exact free set are disjoint. Summing their contributions gives the finite-$t$ identity
\[
 t^{-r}\E T_r(\eta_t)
 =\sum_{|A|=r}\left(\prod_{i\in A}w_i\right)
             \left(\prod_{j\notin A}(1-tw_j)\right)
       \E_{\theta^{(t)}}|\partial_A F(X)|,
\]
where only the complementary coordinates matter inside the expectation. For each fixed finite $m$, this is a finite sum. As $t$ decreases to zero, $\theta_j^{(t)}\to\theta_j$ and the factors $1-tw_j$ tend to one, so we may take the limit termwise. Using the preceding bound on $\E T_r$ yields
\[
 \lim_{t\downarrow0}t^{-r}\E T_r(\eta_t)
 =\sum_{|A|=r}\left(\prod_{i\in A}w_i\right)
       \E_\theta|\partial_A F(X)|\le\Lambda^r/2,
\]
proving \eqref{eq:weighted-mean}. Multilinearity gives $\partial_A F(\theta)=\E_\theta\partial_A F(X)$, so Jensen proves \eqref{eq:weighted-point}. Zero weights require no division, including at the cube boundary. Every deterministic restriction of $C$ satisfies Definition~\ref{def:criticality} with the same $\Lambda$, by composing restrictions; applying the argument to it proves the last assertion.
\end{proof}

\subsection{Exact Gaussian averaging of Boolean tests}\label{subsec:ph-averaging}
Fix a mean vector $\mu\in\R^m$, a noise scale $s>0$, and a vector $Y$ of independent standard normals. The sign $\sgn(\mu_i+sY_i)$ has mean $h_s(\mu_i)=2\Phi(\mu_i/s)-1$. Define
\[
 \Psi_s(\mu)=F(h_s(\mu)),\qquad s>0.
\]
Since the signs are independent, the product-measure interpretation of $F$ gives the exact identity
\begin{equation}\label{eq:exact-averaging}
 \Psi_s(\mu)=\E_Y C(\sgn(\mu+sY)).
\end{equation}
Thus $0\le\Psi_s\le1$. Averaging noise already present in the distribution introduces no approximation error. Both applications below use a fixed positive noise scale.

Derivatives of $h_s(\mu_i)$ are bounded by $w_i=1-|h_s(\mu_i)|$ times a polynomial in $|\mu_i|$, matching Lemma~\ref{lem:weighted} even near $\pm1$. We need the off-diagonal Hessian for covariance interpolation and the third directional derivative for magnitude replacement.

\begin{lemma}[Derivative bounds after Gaussian averaging]\label{lem:ph-smooth}
Suppose $C$ satisfies Definition~\ref{def:criticality}, and let $\Psi_s$ be given by \eqref{eq:exact-averaging}. For fixed $s>0$, $B_\mu=\norm\mu_\infty$, and every direction $v$,
\begin{align}
 \sum_{i\ne j}|\partial_{ij}\Psi_s(\mu)|
   &\le C_s(1+B_\mu)^2\Lambda^2,\label{eq:ph-hessian}\\
 |D_v^3\Psi_s(\mu)|
   &\le C_s\norm v_\infty^3(1+B_\mu)^3
              \Lambda^3.\label{eq:ph-third}
\end{align}
\end{lemma}
\begin{proof}
Write $\theta_i=h_s(\mu_i)$, $w_i=1-|\theta_i|$, and $h_i^{(r)}=h_s^{(r)}(\mu_i)$. Mills' ratio \eqref{eq:mills} and
\[
 h_s''(z)=-\frac z{s^2}h_s'(z),\qquad
 h_s'''(z)=\left(\frac{z^2}{s^4}-\frac1{s^2}\right)h_s'(z)
\]
give $|h_i^{(r)}|\le C_s(1+|\mu_i|)^rw_i$ for $1\le r\le3$. With all derivatives of $F$ evaluated at $\theta$, the chain rule gives
\[
 \partial_{ij}\Psi_s=F_{ij}h_i'h_j'\quad(i\ne j),
\]
\[
 D_v^3\Psi_s=
 \sum_i F_i h_i'''v_i^3+
 3\sum_{i\ne j}F_{ij}h_i''h_j'v_i^2v_j+
 \sum_{\substack{i,j,k\\\text{distinct}}}
       F_{ijk}h_i'h_j'h_k'v_iv_jv_k.
\]
Lemma~\ref{lem:weighted} therefore yields
\begin{align*}
 \sum_{i\ne j}|\partial_{ij}\Psi_s|
 &\le C_s(1+B_\mu)^2\Lambda^2,\\
 |D_v^3\Psi_s|
 &\le C_s\norm v_\infty^3(1+B_\mu)^3
              (\Lambda+\Lambda^2+\Lambda^3)\\
 &\le C'_s\norm v_\infty^3(1+B_\mu)^3\Lambda^3,
\end{align*}
since $\Lambda\ge1$. For the remaining derivatives in covariance interpolation, $\partial_i\Psi_s=F_i h_i'$ and $\partial_{ii}\Psi_s=F_i h_i''$ give
\[
 \sum_i|\partial_i\Psi_s|\le C_s(1+B_\mu)\Lambda,
 \qquad
 \sum_i|\partial_{ii}\Psi_s|\le C_s(1+B_\mu)^2\Lambda.
\]
\end{proof}

\subsection{\texorpdfstring{Comparing $\Pcal_{n,\sigma}$, $\G_{n,\sigma}$, and $\Unif_n$}{Comparing P(n,sigma), G(n,sigma), and U(n)}}
For the Gaussian-sign comparison, retain a fixed independent Gaussian component for exact averaging and interpolate the remaining covariance. A positive eigenvalue lower bound makes this possible throughout the path. The covariance diagonal stays fixed, so Price's formula uses only the off-diagonal Hessian.

\begin{proposition}[Gaussian-sign comparison for Boolean tests]\label{prop:ph-gaussian}
Let $\Sigma$ be a covariance matrix with unit diagonal and smallest eigenvalue at least $\kappa>0$, and set $\beta=\max_{i\ne j}|\Sigma_{ij}|$. Let $X_\Sigma\sim\Normal(0,\Sigma)$, and let $U$ be uniform on $\{\pm1\}^m$. For every Boolean $C$ satisfying Definition~\ref{def:criticality},
\[
 |\E C(\sgn X_\Sigma)-\E C(U)|
 \le C_\kappa\beta\Lambda^2\log(2m).
\]
\end{proposition}
\begin{proof}
Put $s_0^2=\kappa/2$ and $\Gamma_t=(1-t)I+t\Sigma-s_0^2I$. Since the unit diagonal implies $\kappa\le1$,
\[
 \lambda_{\min}(\Gamma_t)
 \ge(1-t)+t\kappa-\kappa/2\ge\kappa/2,
 \qquad (\Gamma_t)_{ii}=1-s_0^2\le1.
\]
For $V_t\sim\Normal(0,\Gamma_t)$, exact averaging \eqref{eq:exact-averaging} over an independent Gaussian component of variance $s_0^2$ identifies the desired difference with $\E\Psi_{s_0}(V_1)-\E\Psi_{s_0}(V_0)$. Price's formula (Lemma~\ref{lem:price}) and Lemma~\ref{lem:ph-smooth} give
\begin{align*}
 \left|\frac d{dt}\E\Psi_{s_0}(V_t)\right|
 &=\left|\frac12\sum_{i\ne j}\Sigma_{ij}
                    \E\partial_{ij}\Psi_{s_0}(V_t)\right|\\
 &\le C_\kappa\beta\Lambda^2
                \E(1+\norm{V_t}_\infty)^2\\
 &\le C'_\kappa\beta\Lambda^2\log(2m).
\end{align*}
The marginal variance bound allows us to apply Lemma~\ref{lem:maxmoment} in the last step. Integrating over $t\in[0,1]$ proves the claim.
\end{proof}
For the covariance $\Sigma_{n,\sigma}$ of our construction, the smallest eigenvalue $1-\rho$ is fixed and the largest off-diagonal entry is $\rho D^{-1/2}$. Taking $m=2D$ gives
\begin{equation}\label{eq:ph-g-uniform}
 |\E_{\G_{n,\sigma}}C-\E_{\Unif_n}C|
 =O(D^{-1/2}\Lambda^2\log(2D)).
\end{equation}

\paragraph{\texorpdfstring{Comparing $\Pcal_{n,\sigma}$ with $\G_{n,\sigma}$.}{Comparing P(n,sigma) with G(n,sigma).}}
We form the Gaussian average \eqref{eq:exact-averaging} with $f$ fixed and apply Lemma~\ref{lem:magnitude}. Its third-derivative bound grows cubically with the largest coordinate of the mean, giving the $k=3$ case of that lemma.

\begin{proposition}[Boolean magnitude replacement]\label{prop:ph-replace}
Let $C:\{\pm1\}^{2D}\to\{0,1\}$ satisfy Definition~\ref{def:criticality}. Then, for each $\sigma\in\{+1,-1\}$,
\begin{equation}\label{eq:ph-replace}
 |\E_{\G_{n,\sigma}}C-\E_{\Pcal_{n,\sigma}}C|
 \le C D^{-1/2}\Lambda^3
                   (\log(2D))^{3/2}.
\end{equation}
\end{proposition}
\begin{proof}
For each $f$, define
\[
 \Psi_f(\mu)=\E_Y C(f,\sgn(\mu+sY)).
\]
Fixing $f$ preserves the criticality bound $\Lambda$. Thus \eqref{eq:exact-averaging} and Lemma~\ref{lem:ph-smooth} give, uniformly in $f$,
\[
 |D_v^3\Psi_f(\mu)|
 \le C_s\Lambda^3\norm v_\infty^3(1+\norm\mu_\infty)^3.
\]
This is \eqref{eq:magnitude-hyp} with $k=3$ and $B=C_s\Lambda^3$, where $s$ is the fixed constant from \eqref{eq:constants}. Lemma~\ref{lem:magnitude} now compares the expectations averaged over uniform $f$. At the fully replaced endpoint, $\rho^2v_0+s^2=b^2$, giving the conditional product distribution $\Pcal_{n,\sigma}$. The original endpoint is the Gaussian-sign reference distribution $\G_{n,\sigma}$. This proves \eqref{eq:ph-replace}.
\end{proof}

\begin{proof}[Proof of Theorem~\ref{thm:finite-common}, Boolean case]
The triangle inequality along the three comparisons gives, for each $\sigma$,
\[
 \begin{aligned}
 |\E_{\Ecal_{n,\sigma}}C-\E_{\Unif_n}C|
 \le{}& O(\Lambda/M+M^{-2})
       +O(D^{-1/2}\Lambda^3(\log(2D))^{3/2})\\
      &{}+O(D^{-1/2}\Lambda^2\log(2D)),
 \end{aligned}
\]
by Proposition~\ref{prop:repair}, Proposition~\ref{prop:ph-replace}, and \eqref{eq:ph-g-uniform}, respectively. The sum of these bounds for $\sigma=+1,-1$ bounds $|\E_{\Ecal_{n,+}}C-\E_{\Ecal_{n,-}}C|$, giving
\begin{equation}\label{eq:ph-natural}
 O\!\left(D^{-1/2}\Lambda^3(\log(2D))^{3/2}
                  +\frac\Lambda M+\frac1{M^2}\right).
\end{equation}
Here $\Lambda\ge1$ absorbs the covariance cost. Since $M=\Theta(D^{1/12})$, for any fixed polynomial $q$, uniformly over $1\le\Lambda\le q(n)$,
\[
 \frac{D^{-1/2}\Lambda^3(\log(2D))^{3/2}}{\Lambda/M}
 =O\!\left(MD^{-1/2}\Lambda^2n^{3/2}\right)=o(1).
\]
Thus the Boolean argument gives the stronger bound $O(\Lambda/M+M^{-2})=O(q(n)/M)$, with count repair determining the leading error. This implies family (ii) of Theorem~\ref{thm:finite-common}, since $q\ge n+1$ and $M=\Theta(2^{n/12})$.

Lemmas~\ref{lem:ph-compile} and~\ref{lem:critical-envelope} place every fixed $\PH$ resource bound in this polynomial envelope, independently of the side-bit values.
\end{proof}
Theorem~\ref{thm:finite-common} now holds for both test families.

\section{Constructing the recursive oracle}\label{sec:oracle}

A \emph{requirement} is a padded public-coin verifier in the normal form of Section~\ref{subsec:full-ip}, or a fixed-alternation predicate, each with an explicit polynomial clock. Lemmas~\ref{lem:am-compile}, \ref{lem:ph-compile}, and~\ref{lem:critical-envelope} give polynomial resource bounds uniform in the side oracle. By Theorem~\ref{thm:finite-common} and Section~\ref{subsec:finite-bound}, every sufficiently large eligible length has a promised pair violating the requirement's completeness or soundness threshold. We find one by exhaustive search and commit all accessible bits to preserve the failure.

\begin{proposition}[Effective oracle diagonalization]
\label{prop:extension}
Let $\mathcal N$ be an infinite set of address lengths decidable in polynomial time on unary input, with the target-block encoding of Section~\ref{subsec:models}. Suppose the following conditions hold.
\begin{enumerate}[label=(\roman*)]
\item At each length, the positive and negative promise sets lie in an explicitly enumerable finite space of target truth tables and are uniformly decidable.
\item Positive default inputs are uniformly computable, and a uniform polynomial-time quantum algorithm decides the sign exactly on promised inputs.
\item The classical requirements are effectively enumerable, each with an explicit polynomial clock. On a fixed input, once every accessible oracle bit is fixed, each requirement is represented by a finite game with Boolean terminal values and uniform-average, maximum, or minimum gates.
\end{enumerate}
Denote the game value by $a_i^O(n)\in[0,1]$. Assume that for each fixed requirement, at every sufficiently large eligible length and uniformly over all accessible side-oracle bits, some promised pair satisfies
\begin{equation}\label{eq:common-failure}
 a_i^O(n)<2/3\quad\text{on a positive pair},\qquad\text{or}\qquad
 a_i^O(n)>1/3\quad\text{on a negative pair}.
\end{equation}
Then there exist a recursive Boolean oracle $O$ and a unary language $L\in\EQP^O$ for which every enumerated requirement fails.
\end{proposition}

For a PH predicate, maxima and minima represent existential and universal choices, so the game value is zero or one. Then \eqref{eq:common-failure} means a wrong answer. For a public-coin verifier, averages represent fresh coins and maxima range over prover replies, so the value is its optimal acceptance probability. The notation $a_i^O(n)$ includes the candidate target pair as part of the provisional oracle $O$.

\begin{proof}
We enumerate program codes with polynomial clocks bounding time, message and coin lengths, and oracle query lengths on every history, including malformed replies or oracle inputs. Every polynomial-resource machine appears with a sufficiently large clock; recognizing polynomial-time programs is unnecessary. In an interactive requirement only the verifier is clocked; the prover's computation and oracle access remain unrestricted. All enumeration orders are deterministic. Initially the empty-string bit is zero and $B_0=0$.

\smallskip\noindent\textbf{Stage $i$.}
Before stage $i$, every string of length at most $B_{i-1}$ has a permanent value, and every eligible block in this prefix satisfies the promise. Try eligible address lengths $n>B_{i-1}$ in increasing order and set
\[
 B=\max\{B_{i-1},n+1,T_i(n)\},
\]
where $T_i$ is the requirement's global clock. The target block uses strings of length $n+1$, including the tag, so it is fresh. The bound $B$ covers both this entire block and every query that the verifier or predicate could make on $1^n$, on any history. Provisionally fill every other new eligible block through $B$ with its positive default input, and set unused strings to zero, preserving the old prefix. These assignments are provisional until a failing target pair is found.

Each address block of length $m$ consists entirely of strings of length $m+1$, so no block is partially committed.

We enumerate all target pairs and test the promised ones. In our application, the $2^{2D}$ pairs of $D$-entry sign tables give a known endpoint even if no pair fails. Every oracle answer needed by the finite game is specified. Backward induction through uniform-average and max/min gates computes its value exactly as a rational number, making both strict threshold tests in \eqref{eq:common-failure} decidable. For verifiers, the maximization covers all legal reply strategies without enumerating prover programs.

When a failure is found, commit the target and provisional side bits, set $B_i=B$, and end the stage. If no promised pair fails at this length, discard its provisional assignments and try the next length. Earlier committed bits never change.

\medskip\noindent\textbf{Termination.}
At each candidate length, the search is finite and decidable. Uniform eventual failure guarantees success at some length, regardless of its provisional side assignment. In our application, the distributions prove termination; the search uses only finite truth tables and exact rational game values, without computing distributional probabilities.

\medskip\noindent\textbf{Preservation.}
Every oracle query made by the clocked classical computation is fixed, so every completion induces the same finite game. An unrestricted prover may use later oracle bits to choose a strategy, but all such reply strategies are already included in its maxima. An optimizer can hardcode the finite response table. Hence the optimal acceptance value, and likewise the PH max/min value, is unchanged by later stages.

\medskip\noindent\textbf{Recursiveness and the separating language.}
The bounds grow strictly, since $B_i\ge n_i+1>B_{i-1}$, where $n_i$ is the successful address length at stage $i$. To compute $O(z)$, simulate the deterministic construction until $B_i\ge|z|$, then return the assigned bit. Every stage terminates and the bounds are unbounded, so this algorithm halts on every input. The oracle is therefore recursive. No running-time bound is asserted.

Every eligible block satisfies the promise, having been committed as a target or positive default input. Let $L$ consist of the eligible unary strings whose block is positive. The quantum algorithm rejects non-unary inputs and lengths outside $\mathcal N$, and otherwise applies the uniform exact solver to the block. It queries the completed oracle without simulating the exhaustive construction. Thus $L\in\EQP^O$, and each classical requirement fails at its committed target input.
\end{proof}

\begin{proof}[Proof of Theorem~\ref{thm:main}]
We interleave the public-coin verifier and fixed-alternation predicate lists so that every candidate receives a stage. Theorem~\ref{thm:finite-common} gives \eqref{eq:common-failure} for each requirement, uniformly over its side assignment. Both families use the same promise sets, uniformly computable default inputs \eqref{eq:filler}, and exact solver from Proposition~\ref{prop:exact-solver}. Membership in the promise sets is decidable by the integer comparison in Section~\ref{subsec:notation}, and eligibility $n\in4\N_{>0}$ is polynomial-time decidable. Proposition~\ref{prop:extension} therefore constructs a recursive oracle and a unary language in $\EQP^O$ defeating both lists.

A private-coin proof for $L$ relative to $O$ would, by Theorem~\ref{ext:gs} and the normal form of Section~\ref{subsec:full-ip}, yield a verifier on the first list, contradicting its failure. The second list excludes every PH predicate. Thus $L\in\EQP^O$, $L\notin\IP^O$, and $L\notin\PH^O$. The quantum machine rejects non-unary and ineligible inputs without queries and uses the three-query solver at every eligible length, including those filled rather than targeted.
\end{proof}

\appendix

\section{Exact implementation with real gates}\label{app:real}

Section~\ref{sec:quantum} gives an exact three-query solver over Clifford+$T$. Here we implement it using only Hadamard and reversible classical gates. The construction uses the quadratic sign $f_\star$ from the default inputs and requires an even number of address bits, as ensured by our eligible lengths.

Let $J_{\rm swap}$ exchange the address halves and let $(P_{\rm swap}g)(x)=g(J_{\rm swap}x)$. Then $P_{\rm swap}^T=P_{\rm swap}=P_{\rm swap}^{-1}$, and $HP_{\rm swap}=P_{\rm swap}H$ because $x\cdot J_{\rm swap}y=(J_{\rm swap}x)\cdot y$. Let $D_{f_\star}$ be the diagonal sign matrix of $f_\star$, implementable without an oracle query.

\begin{lemma}[Quadratic-sign identity]\label{lem:bent-identity}
\[
 HD_{f_\star}H=D_{f_\star}HP_{\rm swap}D_{f_\star}.
\]
\end{lemma}
\begin{proof}
Using $Hf_\star=f_\star$ from Section~\ref{sec:quantum}, with addition in $\mathbb F_2^n$,
\[
 \begin{aligned}
 (HD_{f_\star}H)_{xy}
 &=D^{-1/2}(Hf_\star)(x+y)=D^{-1/2}f_\star(x+y)\\
 &=D^{-1/2}f_\star(x)f_\star(y)(-1)^{x\cdot J_{\rm swap}y}\\
 &=(D_{f_\star}HP_{\rm swap}D_{f_\star})_{xy}.
 \end{aligned}
\]
The second line is the polarization identity for the quadratic exponent $x_L\cdot x_R$.
\end{proof}

\paragraph{Branch preparation.}
For the one-query preparation, the address register starts in a uniform superposition, the branch qubit $b$ in $\ket{+}$, and a copied-address register $c$ in zero. We copy the address into $c$ with CNOTs and swap the halves of $c$ when $b=1$, giving the pre-query state
\[
 \frac1{\sqrt{2D}}\sum_{b\in\{0,1\}}\sum_x
       \ket b\ket x\ket{J_{\rm swap}^{\,b}x}.
\]
Applying $U_O$ to the tagged register $bc$ supplies phase $f(x)$ on branch zero and $g(J_{\rm swap}x)$ on branch one. Uncomputing the swap and copy produces
\[
 \frac1{\sqrt2}\left(
    \ket0\,f/\sqrt D+\ket1\,P_{\rm swap}g/\sqrt D\right)
\]
with the copied register and all auxiliary workspace restored to zero. Here $f/\sqrt D$ and $g/\sqrt D$ denote normalized sign-amplitude vectors. Only known reversible gates need to be controlled, and each controlled swap is a CNOT--Toffoli--CNOT circuit.

\paragraph{The Forrelation overlap.}
Lemma~\ref{lem:bent-identity} lets us replace the controlled Hadamard by unconditional Hadamards and a controlled quadratic phase. We apply $D_{f_\star}$ followed by $H$ to both branches, then $D_{f_\star}$ only to branch zero, and finally $H$ to both branches. This gives
\[
 \begin{aligned}
 \text{branch }0:\quad&
 HD_{f_\star}HD_{f_\star} f/\sqrt D
       =D_{f_\star}HP_{\rm swap}f/\sqrt D,\\
 \text{branch }1:\quad&
 HHD_{f_\star}P_{\rm swap}g/\sqrt D
       =D_{f_\star}P_{\rm swap}g/\sqrt D.
 \end{aligned}
\]
Their overlap is
\[
 \begin{aligned}
 \frac1D(D_{f_\star}HP_{\rm swap}f)^T
                    (D_{f_\star}P_{\rm swap}g)
 &=\frac1D f^TP_{\rm swap}HP_{\rm swap}g\\
 &=\frac1D f^THg=\forr(f,g).
 \end{aligned}
\]
A final Hadamard on the branch qubit therefore yields $b=0$ with probability $(1+\forr(f,g))/2$.

The gate $D_{f_\star}$ applies CZ to each pair $(x_{L,j},x_{R,j})$, and its branch-controlled version uses CCZ with a negative control on $b$. Conjugating a target of CNOT or Toffoli by $H_2$ implements CZ or CCZ, respectively, and $X$ gates implement negative controls. Thus all gates are real.

Equations~\eqref{eq:amp-state}--\eqref{eq:amp-prob} give an exact three-query algorithm. The all-zero reflection on every register of the one-query unitary (branch, address, copied address, and clean workspace) uses reversible conjunctions and $Z=H_2XH_2$, followed by uncomputation. This gives a uniform polynomial-size implementation over the fixed real basis $\{H_2,X,\mathrm{CNOT},\mathrm{Toffoli}\}$ and the oracle phase query.

\section{A criticality bound for regular formulas}\label{app:criticality}

We give the version of Rossman's criticality bound~\cite[Theorem~3]{Rossman19} needed for Lemma~\ref{ext:circuit-criticality}, correcting three steps of its proof while preserving the logarithmic exponent. Following his convention, a formula is \emph{regular} if all gates at each height $i\ge2$ have a common fanin $m_i$; bottom-gate fanins may vary. A depth-$(h+1)$ formula then has $s_{\rm reg}=\prod_{i=2}^{h+1}m_i$ depth-one subformulas.

\begin{lemma}\label{lem:criticality-correction}
A regular AND/OR formula $F$ of depth $h+1$, $h\ge1$, with $s_{\rm reg}\ge1$ depth-one subformulas satisfies $\Pp_{\rho\sim\Res_p}[\DT(F|_\rho)\ge r]\le(p\Lambda)^r$ for $p\in[0,1]$ and integers $r\ge1$, where
\[
 \Lambda=8\cdot60^h\left(1+\frac{\ln s_{\rm reg}}h\right)^h.
\]
\end{lemma}
\begin{proof}
\textit{Jensen normalization.}
For the Jensen step in~\cite[Lemma~12]{Rossman19}, let $\mu_i\ge0$, $a=\sum_i\mu_i\le1$, $t_i,s\ge1$, and $B\ge\max\{1,\sum_i\mu_i e^{t_i}\}$. When $a>0$, normalized Jensen for the concave function $x\mapsto(1+\ln x/s)^s$ on $[1,\infty)$ gives
\[
 \sum_i\mu_i(t_i/s)^s
 \le a\left(1+\frac{\ln(B/a)}s\right)^s
 \le\left(1+\frac{\ln B}s\right)^s.
\]
The middle expression is increasing in $a$, since its derivative is $x(1+x)^{s-1}\ge0$ for $x=\ln(B/a)/s\ge0$. The case $a=0$ is immediate. In Propositions~25 and~27 the bound $B=m^s\ge1$ is already available, so those estimates remain unchanged.

\textit{Serial exposure.}
A canonical branch can terminate while other variables remain free, contrary to the assumption in the counting argument of~\cite[Lemma~28]{Rossman19}. We instead reveal coordinates only as the scan reaches them. Use $\{0,1\}$ notation and fix ordered depth-one formulas $F_1,\ldots,F_k$, lengths $s_j\ge1$, answer strings $a_j\in\{0,1\}^{s_j}$, and overwrite strings $c_j\in\{0,1\}^{s_1+\cdots+s_{j-1}}$. Before stage $j$, previously queried variables receive the bits of $c_j$ in concatenated query order. We bound the complete-branch event by the following larger event.

We sample $\rho$ with independent coordinate probabilities $p,(1-p)/2,(1-p)/2$ for $*,0,1$. At each stage we scan the literal/constant list in the order of~\cite[Definition~19]{Rossman19}, revealing a coordinate's state only when first encountered. An OR scan stops at a true literal and skips false literals; the roles reverse for AND. A newly revealed star is queried using the next bit of $a_j$. Its answer follows the same rule, except that we declare success immediately after the $s_j$-th query, without inspecting the suffix or requiring termination there. If the scan stops or runs out of literals earlier, the stage fails. Before the next stage, queried stars receive the prescribed overwrite values, while revealed fixed coordinates retain their $\rho$-values. Constants and repeated literals of either polarity are evaluated from these known values.

On the original joint event, each scan follows its canonical branch through the prescribed queries, with the same query order and overwrites. Thus every stage succeeds. Conditional on the transcript so far, unseen coordinates remain independent with their original laws. Each new coordinate is free with probability $p$, or fixed so that its literal continues or stops the scan with probability $(1-p)/2$ each. Previously revealed values can only be skipped or stop the scan. The conditional probability of reaching another query is therefore at most
\[
 p\sum_{\ell\ge0}\left(\frac{1-p}{2}\right)^\ell
 =\frac{2p}{1+p}\le2p.
\]
Since every exposed star is queried and assigned, the same bound holds at each stage. Conditioning successively on the revealed values gives $(2p/(1+p))^{\sum_j s_j}\le(2p)^{\sum_j s_j}$, as required.

\Needspace{10\baselineskip}
\textit{Induction and restriction density.}
Put $\lambda=\prod_{i=2}^{h+1}8e^2(1+\ln m_i)$ and take $0\le u\lambda\le1$. The successive densities in~\cite[proof of Theorem~3]{Rossman19} have product $u/8$, rather than $u$:
\[
 \frac1{4e}
 \prod_{i=2}^{h}\frac1{8e^2(1+\ln m_i)}
 \frac{u\lambda}{16e(1+\ln m_{h+1})}
 =\frac u8.
\]
At the initial density $1/(4e)$, serial exposure gives the joint bound $(1/(2e))^{s_1+\cdots+s_k}$ for prescribed answer and overwrite strings. Proposition~27 preserves this joint bound for depth-$i$ subformulas at each intermediate density $1/[8e^2(1+\ln m_i)]$, $2\le i\le h$; the Jensen correction leaves its estimate unchanged. At the final density $u\lambda/[16e(1+\ln m_{h+1})]$, Proposition~25 bounds the probability of a complete branch with a fixed answer string of length $s$ by $(u\lambda/4)^s$. These are the induction statements in~\cite[proof of Theorem~3]{Rossman19}, now supplied by the corrected joint branch estimate.

If decision-tree depth is at least $r$, some complete canonical branch has length $s\ge r$. For $0\le u\lambda\le1$, a union bound over answer strings and lengths gives
\[
 \sum_{s\ge r}2^s(u\lambda/4)^s
 =\frac{(u\lambda/2)^r}{1-u\lambda/2}
 \le(u\lambda)^r.
\]
The composed restriction has density $u/8$. Taking $u=8p$ therefore yields $(8p\lambda)^r$ for $p\le1/(8\lambda)$; for larger $p$ the bound is trivial. AM--GM and $8e^2<60$ give the stated $\Lambda$.
\end{proof}

\Needspace{9\baselineskip}
\section*{AI Disclosure}
\phantomsection
\addcontentsline{toc}{section}{AI Disclosure}
\label{sec:ai-methodology}

Our initial idea was to start from a distribution already known to be hard to distinguish from uniform, transfer this hardness to one approximating the desired exact promise, and then apply a suitable cutoff or repair step to obtain exact support. We developed this strategy through several rounds of interaction with ChatGPT (6 Astra and 6.1 Sol), first exploring constructions for extremal $k$-Forrelation with promise values $\pm1$, for both $k=2$ and higher $k$.

For $k=2$, changing the promise values to $\pm1/2$ led to the count-repair approach. ChatGPT 6.1 Sol generated the resulting construction and its proof, which we verified, simplified, and further developed.

AI assistance contributed substantially to the construction and comparison of the distributions in Sections~\ref{sec:laws}--\ref{sec:ph}, and to the development of the oracle-separation argument in Section~\ref{sec:oracle}. ChatGPT also assisted with the organization and exposition of the manuscript. We take full responsibility for the correctness, originality, and integrity of the manuscript, including its proofs and references.

\end{document}